\documentclass[final,3p,times]{elsarticle}

\usepackage{amsmath,hyperref}
\usepackage{lineno}

\usepackage{epstopdf}
\journal{Journal}

\usepackage{lineno,hyperref}
\usepackage{graphicx}
\usepackage{dcolumn}
\usepackage{bm}
\usepackage{epsfig}
\usepackage{booktabs}
\usepackage{subfigure}
\usepackage{graphics}
\usepackage{amssymb}
\usepackage{amsmath}
\usepackage{array}
\usepackage{color}
\usepackage{booktabs}
\usepackage{multirow}
\usepackage{caption}
\usepackage{chngpage}
\usepackage{subfigure}
\usepackage{mathrsfs,gensymb,float}
\usepackage[table,xcdraw]{xcolor}

\biboptions{numbers,sort&compress}
\modulolinenumbers[1]
\begin{document}

\captionsetup[figure]{labelfont={bf},name={Fig.},labelsep=period}        

\begin{frontmatter}
	
	\title{A diffuse-interface model for \textit{N}-phase flows with liquid-solid phase change}


	\author[1]{Jiangxu Huang}
	\author[1]{Chengjie Zhan}
	\author[1,3,4,5]{Zhenhua Chai\corref{mycorrespondingauthor}}	
	\ead{hustczh@hust.edu.cn}
	\cortext[mycorrespondingauthor]{Corresponding author}
	\author[1]{Changsheng Huang}	
	\author[1]{Xi Liu}
	
	\address[1]{ School of Mathematics and Statistics, Huazhong University of Science and Technology, Wuhan 430074, China}
	\address[3]{Institute of Interdisciplinary Research for Mathematics and Applied Science,Huazhong University of Science and Technology, Wuhan 430074, China}	
	\address[4]{Hubei Key Laboratory of Engineering Modeling and Scientific Computing, Huazhong University of Science and Technology, Wuhan 430074, China}	
	\address[5]{The State Key Laboratory of Intelligent Manufacturing Equipment and Technology, Huazhong University of Science and Technology, Wuhan 430074, China}


\begin{abstract}
In this work, we first propose a diffuse-interface model for simulating \textit{N}-phase flows with solid–liquid phase change. In this model, a phase-field approach is adopted to capture multiphase fluid interfaces, and an enthalpy-based formulation is used to describe the phase change. The volume changes resulting from density differences during phase change are incorporated by introducing a source term into the continuity equation. The method also satisfies the reduction-consistent property, allowing it to rigorously degenerate to both the conservative phase-field method for \textit{N}-phase flows and the classical enthalpy method for solid–liquid phase change. Then a coupled lattice Boltzmann (LB) method is developed to solve this diffuse-interface model. Some numerical tests, including film freezing, single droplet freezing, and compound droplet freezing are performed, and the results are in good agreement with the analytical solutions and data reported in the previous works. Furthermore, the proposed method is applied to study freezing dynamics of complex systems with insoluble impurities, capturing the interaction between the advancing freezing front and embedded impurities. It is found that the proposed diffuse-interface method is accurate and efficient for studying \textit{N}-phase systems with phase change.

\end{abstract}

	
\end{frontmatter}

\section{Introduction}

The solidification of \textit{N}-phase fluids, composed of two or more liquid phases, is a fundamental phenomenon that has been widely observed in both natural environments and industrial applications, such as the dynamics of sea-ice layers \cite{HuerreARFM2024,DuNRP2024}, aerospace engineering \cite{LynchPAS2001}, additive manufacturing \cite{ZhangNC2020}, and the fabrication of electronic devices \cite{LimAFM2008}. This process involves rich interfacial phenomena, the interaction between different phases, and the coupling of fluid flow and heat transfer, bringing some significant challenges in the study of multiphase fluid solidification. To efficiently and accurately control the solidification of multiphase fluids and advance related technologies, it is crucial to gain a comprehensive understanding of the freezing dynamics of multiphase fluids.

The freezing behavior of a pure liquid droplet has been extensively investigated theoretically and experimentally, providing some significant insights into this complex phenomenon, for instance, the propagation of the freezing front \cite{ZhangATE2017,TembelyJFM2019}, contact line pinning mechanisms \cite{DePRF2017,HerbautPRF2019,ShikhmurzaevPOF2021,KoldeweijPRF2021}, droplet spreading during solidification \cite{ZadrazilJFM2006,ThievenazJFM2019,GrivetPRF2023,KantPNAS2020}, and the formation of singular conical tips \cite{MarinPRL2014,ZhangATE2019,StarostinJCIS2022,MiaoATE2024}. In terms of front propagation, Zhang et al. \cite{ZhangATE2017} and Tembely et al. \cite{TembelyJFM2019} proposed theoretical models to predict the evolution of the freezing front and the total freezing time. Regarding contact line dynamics, de Ruiter et al. \cite{DePRF2017} demonstrated that contact line motion is thermally arrested at a critical temperature. Once spreading couples with phase change, the macroscopic hydrodynamics become highly sensitive to thermophysical parameters. Specifically, Zadrazil et al. \cite{ZadrazilJFM2006} detailed the roles of the Stefan number and thermal conductivity during solidification on porous media. In parallel, researchers have quantified the dependence of the solidification rate on local temperature gradients \cite{KantPNAS2020} and evaluated the thickness of the residual solid layer formed during the spreading phase \cite{ThievenazJFM2019}. In addition, the formation of the singular tip has also received increasing attention. For example, Marin et al. \cite{MarinPRL2014} first noted the self-similar conical tip formation, and found that it is independent of the substrate temperature and contact angle. Zhang et al. \cite{ZhangATE2019} illustrated that a sudden change in the gas–liquid interface angle is attributed to the formation of a kinked tip, while Starostin et al. \cite{StarostinJCIS2022} demonstrated that the tip orientation is governed by the direction of heat flux, Miao et al. \cite{MiaoATE2024} reported that the forced convection can regulate tip singularity. Finally, some works have experimentally investigated the effects of external factors (e.g., gravity \cite{ZengPRF2022}, magnetic fields \cite{MenglongIJMF2021}, electric fields \cite{HuangIJHMT2025}, and nanoparticles \cite{ZhaoAPL2021}) on the freezing process.

In the past years, the numerical simulation has also emerged as a powerful tool for understanding the dynamics of single-phase droplet freezing. The primary challenge in simulating the freeing process is the simultaneous tracking of two dynamic interfaces: the solid–liquid phase change front and the gas–liquid interface. Thus the mathematical models for droplet freezing usually adopt two interface-capturing techniques, i.e., the sharp-interface and diffuse-interface methods. Based on the combination of these two different type approaches, the numerical methods for droplet freezing can be classified into three main categories, including multiple sharp interface method \cite{VuIJMF2015,ShetabivashJCP2020}, the multiple diffuse-interface method \cite{ZhangPRE2020,HuangJCP2022, ZhangJCP2022, ZhangJCP2024, ZhangJCP2025,MohammadipourJFM2024, HuangPRE2024, HuangIJHMT2025} and the hybrid diffuse-sharp interface method \cite{LyuJCP2021,ThirumalaisamyIJMF2023,WeiJCP2025,YeCAMC2024}. In summary, the development of these numerical approaches has significantly advanced our understanding on the droplet solidification. However, it should be noted that most of the previous works mainly focus on the solidification behavior of pure liquid droplet, whereas the multiphase fluids are far more representative of real-world industrial conditions. Due to complex physicochemical interactions between different constituents, these  multicomponent fluids exhibit the distinct solidification characteristics that are different significantly from the single-component counterparts. Specifically, the variations in cooling rates can induce the changes in the internal distribution of components, thereby affecting the overall solidification dynamics. Therefore, understanding the solidification dynamics of multiphase fluids is a critical area of research, and some studies have been conducted to explore the complex dynamics. For example, Kant et al. \cite{KantPRL2020} experimentally investigated the freezing of impacting binary droplets on a supercooled surface, finding that the multicomponent nature coupled with evaporation has an important influence on the freezing dynamics of the binary droplets. Meijer et al. \cite{MeijerPRL2023,MeijerSM2024} studied the solidification of water-in-oil droplets dispersed in water, discovering that the freezing of such hierarchical emulsions can trigger topological transitions and induce unexpected local micro-structural changes. Similarly, van Buuren et al. \cite{BuurenPRL2024} found that the freezing front would suffer from significant deformation when it encounters insoluble droplets or bubbles, and it is mainly affected by interfacial dynamics and the thermocapillary effect. Chu et al. \cite{ChuPRF2019} experimentally investigated bubble precipitation during droplet freezing, and demonstrated that the formation rate is influenced by gas solubility, degree of supercooling, freezing time, droplet size, and surface contact angle.

Although these experimental studies have provided some valuable insights and data regarding the solidification of multiphase fluids, they are often limited in achieving the high-precision measurements of temperature and concentration fields, as well as multiphase interface morphology. On the contrary, the numerical method, as an alternative to experimental approach, can be used to  overcome above limitations. However, the existing numerical works mainly focused on the solidification dynamics of pure fluids, with limited attention paid to the complex interactions between  different species in multiphase systems. This research gap motivates the present work, in which we propose an efficient and versatile numerical method for simulating the solidification behavior of multiphase fluids. In this method, the phase-field and enthalpy methods are coupled to describe the freezing process, where the coupling between heat transfer and phase change, as well as the volume change arising from the solid–liquid density difference are fully incorporated.

The remainder of this paper is structured as follows. In Section \ref{sec2}, we present the mathematical model for the containerless freezing of multiphase fluids, followed by the diffuse-interface LB method in Section \ref{sec3}. In Section \ref{sec4}, the numerical results and discussion are shown, and finally, a brief summary is given in Section \ref{sec5}.

\section{Problem statement and governing equations}
\label{sec2}

\begin{figure}[H]
	\centering
	\includegraphics[scale=0.6]{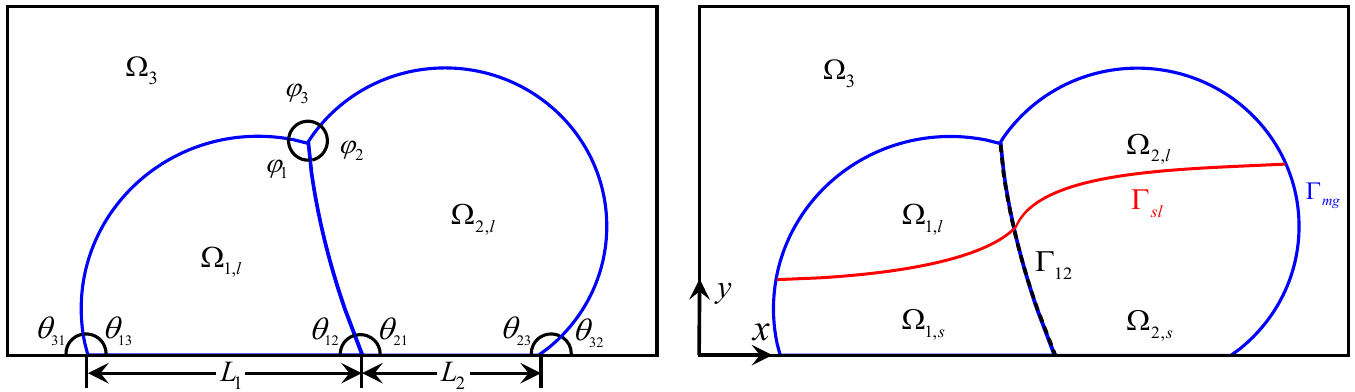}
	\put(-388 ,103){(\textit{a})}
	\put(-188 ,103){(\textit{b})}
	\caption{ Schematic of the compound droplet freezing problem. (a) Initial state of a compound droplet on a solid substrate. (b) Schematic of compound droplet freezing on a cold substrate. The phases involved are as follows: the ambient fluid is denoted by $\Omega_3$ ($\phi_1=0$, $\phi_2=0$, $\phi_3=1$, $f_s=0$), the fluid 1 is represented by $\Omega_{1,l}$ ($\phi_1=1$, $\phi_2=0$, $\phi_3=0$, $f_s=0$), the frozen fluid 1 is denoted by $\Omega_{1,s}$ ($\phi_1=1$, $\phi_2=0$, $\phi_3=0$, $f_s=1$), the fluid 2 is represented by $\Omega_{2,l}$ ($\phi_1=0$, $\phi_2=1$, $\phi_3=0$, $f_s=0$), the frozen fluid 2 is denoted by $\Omega_{2,s}$ ($\phi_1=0$, $\phi_2=1$, $\phi_3=0$, $f_s=1$). $\Gamma_{sl}$ indicates the freezing front between the solid and liquid phases, and $\Gamma_{mg}$ represents the external interface between the compound droplet system and the surrounding ambient fluid.}
	\label{fig1}
\end{figure}

In this section, we will develop a diffuse-interface model for the quasi-equilibrium isotropic freezing process of \textit{N}-phase immiscible fluids without protrusions or dendritic structures in the ambient fluid. Here we take the freezing of a compound droplet on a cold wall as an example, and show the schematic diagram of the problem in Fig. \ref{fig1}. In the rectangular domain $\Omega = \Omega_1 \cup \Omega_2 \cup \Omega_3$, a compound droplet is placed on a cold substrate with an equilibrium contact angle $\theta_{pq}$ ($p,q=1,2,3$, and $p \neq q$), and is surrounded by an ambient fluid. It should be noted that the region $\Omega_p$ is filled with the fluid $p$, and $\theta_{pq}$ represents the pairwise contact angle between the solid wall and the fluid interface formed by phases $p$ and $q$. The interfacial angle $\varphi_p$ define the angle between the phase interfaces, and the value can be used to reflect a balance of surface tensions among different phases. Actually, the interfacial angles satisfy the geometric constraint $\varphi_1 + \varphi_2 + \varphi_3 = 2\pi$ and the surface tension balance condition \cite{ZhangJFM2021}, 
\begin{equation}
	\frac{\sin \varphi_1}{\sigma_{23}} = \frac{\sin \varphi_2}{\sigma_{31}} = \frac{\sin \varphi_3}{\sigma_{12}},
	\label{eq0}
\end{equation}
where $\sigma_{pq}$ is the surface tension coefficient between fluids $p$ and fluids $q$. As the freezing progresses [see Fig. \ref{fig1}(b)], the fluid $p$ solidifies into solid phase $\Omega_{p,s}$, driving continuous advancement of the freezing front $\Gamma_{sl}$. Simultaneously, the volume changes induced by solid-liquid density differences promote the dynamic evolutions of interfaces $\Gamma_{mg}$ and $\Gamma_{12}$, where $\Gamma_{mg}$ denotes the interface between the solid-liquid mixture and the ambient fluid, and $\Gamma_{12}$ represents the contact interface.

In this study, we propose a diffuse-interface model to describe the evolutions of phase interfaces during the freezing process, in which a coupled phase-field and enthalpy method is adopted. The phase-field method is used to capture the dynamics of the fluid-fluid interfaces ($\Gamma_{mg}$), while the enthalpy method is applied for the evolution of the solid-liquid phase change interface ($\Gamma_{sl}$). To this end, a phase-field order parameter $\phi_p$ is introduced to label the interface between the solid-liquid mixture and the residual fluid $q$.  Here $\phi_p=1$ and $\phi_q=0$ ($1 \leq p \neq q \leq \textit{N}$) denotes the solid-liquid mixture of phase $p$, while $\phi_p=0$ and $\phi_q=1$ represents the residual fluid $q$. Additionally, the solid fraction $f_s$ in the enthalpy method is adopted to depict the solid-liquid interface within the solid-liquid mixture, with $f_s=1$ and $f_s=0$ indicating the solid and liquid phases. With the help of the order parameter $\phi_p$ and solid fraction $f_s$, the fluid $p$, frozen fluid $p$, and ambient fluid are represented as ($\phi_p=1$, $\phi_q=0$, $f_s=0$), ($\phi_p=1$, $\phi_q=0$, $f_s=1$), and ($\phi_p=0$, $\phi_N=1$, $f_s=0$), respectively. The \textit{N}-th phase fluid, defined as the non-phase-change ambient fluid, primarily serves as a movable and deformable boundary for the solidification process. In this case, the physical property of the system can be described by a linear function of the order parameter $\phi_p$ and solid fraction $f_s$,
\begin{equation}
	\xi=\sum_{p=1}^{N-1}\left[f_s \phi_p \xi_p^s+\left(1-f_s\right) \phi_p \xi_p^l\right]+\phi_N \xi_N,
	\label{eq1}
\end{equation}
where the parameter $\xi$ denotes the density, viscosity, thermal conductivity, and heat capacity, $\xi_p^l$ and $\xi_p^s$ represent the liquid and solid phases of $p$-th phase fluid, respectively. Under the condition of $\xi_N=\xi_N^s=\xi_N^l$, Eq. (\ref{eq1}) can be rewritten as
\begin{equation}
	\xi=\sum_{p=1}^N\left[f_s \phi_p \xi_p^s+\left(1-f_s\right) \phi_p \xi_p^l\right]. 
	\label{eq2}
\end{equation}

In the following, we present the governing equations for \textit{N}-phase flows with solid-liquid phase change, and only the Newtonian and incompressible fluids are considered. Following the previous works \cite{HuangPRE2024, HuangIJHMT2025}, the momentum equation on the entire domain can be written as 
\begin{equation}
	\frac{\partial \rho \mathbf{u}}{\partial t}+\nabla \cdot(\rho \mathbf{u u})=-\nabla p+\nabla \cdot\left[\mu\left(\nabla \mathbf{u}+(\nabla \mathbf{u})^{\mathrm{T}}\right)\right]+\mathbf{G}+\mathbf{F}_s+\rho \mathbf{f},
	\label{fluid1}
\end{equation}
where $\rho$, $\mathbf{u}$, $p$, $\mu$ are the density, velocity, pressure and dynamic viscosity, respectively. The symbols $\mathbf{G}$ and $\mathbf{F}_s$ represent the body force and the surface tension force. In the presence of solidification, the flow is confined to the fluid regions with a no-slip boundary condition at the solid-fluid interface. This condition is enforced by incorporating a forcing term, $\rho \mathbf{f}$, into the momentum equation, which can be considered as a penalty term that drives the velocity to be zero in the solid region, this technique has been commonly used in the simulations of fluid-solid interaction  problems \cite{HuangPRE2024,HuangIJHMT2025}.

For multiphase flows without phase change, the continuity equation is given by $\nabla \cdot \mathbf{u} = 0$. However, for the problems involving phase change, the condition of a divergence-free velocity field should be modified to account for the volume change induced by the density difference during solid-liquid phase change. Actually, the continuity equation can be expressed as
 \begin{equation}
 	\nabla \cdot \mathbf{u}=-\frac{1}{\rho} \frac{D \rho}{D t},
 	\label{eq5}
 \end{equation}
where $D(\cdot) / D t=\partial_t(\cdot)+\mathbf{u} \cdot \nabla(\cdot)$ is the total derivative. With the help of Eq. (\ref{eq2}), we can rewrite Eq. (\ref{eq5}) as
\begin{equation}
	\begin{aligned}
		& \nabla \cdot \mathbf{u}=-\frac{1}{\rho} \frac{D \rho}{D t}=-\frac{1}{\rho} \sum_p\left[\left(\rho_p^s-\rho_p^l\right) \frac{D f_s \phi_p}{D t}+\rho_p^l \frac{D \phi_p}{D t}\right] \triangleq \dot{m},
		\label{fluid2}
	\end{aligned}
\end{equation}
where the source term $\dot{m}$ on the right-hand side of the continuity equation describes the volume change during the freezing process, and it is influenced by the density ratio of solid to liquid phase.

For the temperature field, the following enthalpy equation is considered \cite{HuangPRE2024,HuangIJHMT2025},
\begin{equation}
	\frac{\partial \left(\rho H\right)}{\partial t}+\nabla \cdot\left(\rho C_p T \mathbf{u}\right)=\nabla \cdot(\lambda \nabla T) + \rho C_p T \dot{m},
	\label{eq4}
\end{equation}
where $T$, $C_p$ and $\lambda$ are the temperature, specific heat capacity and thermal conductivity, respectively. The total enthalpy is defined as $H=C_p T+L f_l $, where $L$ is the latent heat and $f_l$ is the liquid fraction. The last term on the right-hand side of the above equation originates from the source term $\dot{m}$ in the continuity equation, and can be removed when the volume change are neglected. Once the mixture total enthalpy $\rho H$ is known, the temperature $T$ and liquid phase fraction $f_l$ can be determined by \cite{ZhaoAML2020}
\begin{equation}
	f_l =\left\{\begin{array}{ll}
		0 & \rho H< \rho_s H_s \\
		\frac{\rho H - \rho_s H_s}{\rho_l H_l-\rho_s H_s} &\rho_s  H_s \leqslant \rho H \leqslant \rho_l H_l, \\
		1 & \rho H>\rho_l H_l
	\end{array} \quad T= \begin{cases} \frac{\rho H}{\rho_s  C_{p,s}}  & \rho H< \rho_s H_s \\
		T_s+\frac{\rho H-\rho_s H_s}{\rho_l H_l-\rho_s H_s}\left(T_l-T_s\right) & \rho_s H_s \leqslant \rho H \leqslant \rho_l H_l, \\
		T_l+ \frac{\rho H-\rho_l H_l}{\rho_l C_{p,l}}  & \rho H>\rho_l H_l\end{cases}\right.
	\label{H}
\end{equation}
where $T_s$ and $T_l$ are the solidus and liquidus temperatures, $H_s$ and $H_l$ are the corresponding total enthalpies. It is worth noting that the present formulas for calculating $f_l$ and $T$ take into account the solid-liquid density difference, and reduce to the conventional forms when $\rho_s = \rho_l $. In addition, the solid fraction $f_s = 1 - f_l$ is defined as 0 in the liquid phase, 1 in the solid phase, and $0<f_s<1$ in the mushy zone.

The evolution of the order parameter, $\phi_p$ is governed by the conservative Allen-Cahn (AC) equation for immiscible \textit{N}-phase flows, which can be written as \cite{ZhanCICP2024,MirjaliliJCP2024}
\begin{equation}
	\frac{\partial \phi_p}{\partial t}+\nabla \cdot\left(\phi_p \mathbf{u}\right)=\nabla \cdot M\left(\nabla \phi_p-\mathbf{R}_p\right), \quad 1 \leq p \leq N,
	\label{eq_ph}
\end{equation}
where order parameter $\phi_p$ is the volume fraction of phase $p$ in total volume, and satisfies the volume conservation condition $\sum_p \phi_p=1$, $ M $ is the mobility, $\mathbf{R}_p$ is the source term, and is given by \cite{MirjaliliJCP2024}
\begin{equation}
	\mathbf{R}_p=\sum_{q \neq p} \frac{4 \phi_p \phi_q}{\epsilon} \mathbf{n}_{p q},
\end{equation}
here $\mathbf{n}_{p q}=\nabla \phi_{p q} / |\nabla \phi_{p q}|$ ($q \neq p$) is the pairwise normal vector with $\phi_{p q}=\phi_p /\left(\phi_p+\phi_q\right)$ being the pairwise volume fraction. Compared to other \textit{N}-phase phase-field models, the present one is more robust and accurate, and also can be used to avoid spurious numerical artifacts. In addition, it also satisfies the reduction-consistent property, which means that in the absence of $M$ phases, the model would reduce to the one for an $(N-M)$ phase flows. Now, let us focus on the surface tension force in Eq. (\ref{fluid1}). In the phase-field theory, a reduction-consistent total free energy of \textit{N}-phase fluid system can be written as \cite{ZhanCICP2024}
\begin{equation}
	\mathcal{F}(\phi,\nabla \phi)=\int_{\Omega}\sum_{p \neq q} \left[E_0(\phi)+ \frac{k_{p q}}{2} \nabla \phi_p \cdot \nabla \phi_q\right] d \Omega, \quad \text { for } 1 \leq p, q \leq N,
\end{equation}
where $E_0(\phi)=\beta_{p q}\left[g\left(\phi_p\right)+g\left(\phi_q\right)-g\left(\phi_p+\phi_q\right)\right]$ is a bulk free energy with the function $g(\phi)=\phi^2(1-\phi)^2$, $k_{p q}$ and $\beta_{p q}$ are two physical parameters related to the interface thickness $\epsilon$ and  the pairwise symmetric surface tension coefficient $\sigma_{pq}$,
\begin{equation}
	\beta_{p q}=\frac{3}{\epsilon} \sigma_{p q}, \quad k_{p q}=-\frac{3 \epsilon}{4} \sigma_{p q},
\end{equation}
where $\beta_{p q}$ satisfies $\beta_{p q}=\beta_{q p}$ ($p \neq q$) and $\beta_{p p}=0$. Based on the mixing energy mentioned above, one can also derive the chemical potential $\mu_{\phi_p}$,
\begin{equation}
	\mu_{\phi_p}=\frac{\delta \mathcal{F}}{\delta \phi_p}=\sum_{q \neq p} 2 \beta_{p q}\left[g^{\prime}\left(\phi_p\right)-g^{\prime}\left(\phi_p+\phi_q\right)\right]-\sum_{q \neq p} k_{p q} \nabla^2 \phi_q .
\end{equation}
Then the surface tension force can be given by
\begin{equation}
	\mathbf{F}_s=\sum_p \mu_{\phi_p} \nabla \phi_p.
\end{equation}

\section{Numerical methods}
\label{sec3}
In this work, the LB method is used to solve above diffuse-interface model, which is composed of the incompressible Navier-Stokes equations for fluid flows convection-diffusion equations for the phase-field and enthalpy variables. The details on the LB method for solving such coupled systems can be found in Ref. \cite{ChaiPRE2020}, while for brevity and completeness, here we only give a brief overview on this method. 

For the flow field, the LB evolution equation can be expressed as \cite{HuangPRE2024,HuangIJHMT2025}:
\begin{equation}
	f_i\left(\mathbf{x}+\mathbf{c}_i \Delta t, t+\Delta t\right)-f_i(\mathbf{x}, t)=-\frac{1}{\tau_f}\left[f_i(\mathbf{x}, t)-f_i^{\mathrm{eq}}(\mathbf{x}, t)\right]+\Delta t\left(1-\frac{1}{2 \tau_f}\right) F_i(\mathbf{x}, t),
\end{equation}
where $f_i(\mathbf{x}, t)$ is the distribution function for flow field at position $\mathbf{x}$ and time $t$. $\tau_f =\mu/ (\rho c_s^2 \delta t) + 0.5 $ is the relaxation time with $c_s=c/\sqrt{3}$ being the lattice sound speed, $c=\Delta x / \Delta t$ is the lattice speed with $\Delta x$ denoting the lattice spacing. The local equilibrium distribution function $f_i^{\mathrm{eq}}$ is given by
\begin{equation}
	f_i^{\mathrm{eq}}= \begin{cases}\frac{p}{c_s^2}\left(\omega_i-1\right)+\rho s_i(\mathbf{u}), & \mathrm{i}=0 \\ \frac{p}{c_s^2} \omega_i+\rho s_i(\mathbf{u}), & \mathrm{i} \neq 0\end{cases}
\end{equation}
with
\begin{equation}
	s_i(\mathbf{u})=\omega_i\left[\frac{\mathbf{c}_i \cdot \mathbf{u}}{c_s^2}+\frac{\left(\mathbf{c}_i \cdot \mathbf{u}\right)^2}{2 c_s^4}-\frac{\mathbf{u} \cdot \mathbf{u}}{2 c_s^2}\right].
\end{equation}

The force term $F_i$ can be expressed as
\begin{equation}
	F_i = \omega_i \left[ S + \frac{\mathbf{c}_i \cdot (\mathbf{F} + \mathbf{f})}{c_s^2} + \frac{(\mathbf{c}_i\mathbf{c}_i - c_s^2 \mathbf{I}) : \mathbf{\Lambda}}{2c_s^4} \right],
\end{equation}
where $S=\rho \dot{m}+\mathbf{u} \cdot \nabla \rho$, $\mathbf{\Lambda}=\mathbf{u}\mathbf{F} + \mathbf{F}\mathbf{u} + \frac{2}{d} \rho c_s^2 S \mathbf{I}$ with $d$ being dimension, $\mathbf{F}=\mathbf{F}_{\mathrm{s}}+\mathbf{G}$. With the Chapman-Enskog analysis, the Naiver-Stokes equations can be recovered from the LB model, and the velocity $\mathbf{u}$ and pressure $p$ can be calculated by
\begin{subequations}
\begin{equation}
	\rho \mathbf{u}^*=\sum \mathbf{c}_i f_i+\frac{\Delta t}{2} \mathbf{F}, \quad \mathbf{u}=\mathbf{u}^*+\frac{\Delta t}{2} \mathbf{f},
\end{equation}
\begin{equation}
	p=\frac{c_s^2}{\left(1-\omega_0\right)}\left[\sum_{i \neq 0} f_i+\frac{\Delta t}{2} S+(\tau_f-0.5) \Delta t F_0+\rho s_0(\mathbf{u})\right],
\end{equation}
\end{subequations}
here $\mathbf{u}^*$ is the velocity without considering the solid-liquid interaction, the force $\mathbf{f}=f_s(\mathbf{u}_s-\mathbf{u}^*)/\Delta t$ is used to indicate fluid-solid interaction with $\mathbf{u}_s$ being the solid-phase velocity. We note that this approach for dealing with the solid-liquid interaction has been widely used to study the particulate flows \cite{LiuCICP2022, LiuCF2024}, dendrite growth \cite{ZhanCICP2023}, phase change and fluid flows in complex geometries \cite{HuangPRE2024, HuangIJHMT2025, ZhanPD2024}.

For the temperature field, the evolution equation of LB method can be similarly given by \cite{HuangJCP2015}
\begin{equation}
	h_i\left(\mathbf{x}+\mathbf{e}_i \Delta t, t+\Delta t\right)=h_i(\mathbf{x}, t)-\frac{1}{\tau_h}\left[h_i(\mathbf{x}, t)-h_i^{e q}(\mathbf{x}, t)\right]+\left(1-\frac{1}{2 \tau_h}\right) \Delta t H_i,
\end{equation}
where $h_i(\mathbf{x}, t)$ is the distribution function for temperature field, $\tau_h=\lambda/(\rho_{\text{ref}} C_{p, \text { ref }} c_s^2 \Delta t)+0.5$ is the relaxation time. The distribution functions $h_i^{e q}$ and $H_i$ are given by 
\begin{equation}
	h_i^{e q}= \begin{cases}\rho H - \rho C_{p, \text { ref }} T+\omega_i \rho C_p T\left(\frac{C_{p, \text { ref }}}{C_p}-\frac{\mathbf{I}: \mathbf{u u}}{2 c_s^2}\right), & i=0, \\ \omega_i \rho C_p T\left[\frac{C_{p, \text { ref }}}{C_p}+\frac{\mathbf{e}_i \cdot \mathbf{u}}{c_s^2}+\frac{\left(\mathbf{e}_i \mathbf{e}_i-c_s^2 \mathbf{I}\right): \mathbf{u u}}{2 c_s^4}\right], & i \neq 0,\end{cases},\quad H_i= \rho C_p T \nabla \cdot \mathbf{u},
\end{equation}
where $c_{p, \text { ref }}=2 C_{p, s} C_{p, l} /\left(C_{p, s}+C_{p, l}\right)$ is the reference specific heat capacity. Through asymptotic analysis, the enthalpy equation (\ref{eq4}) can be recovered from this LB method, and the total enthalpy is computed by
\begin{equation}
	H=\sum_{i=0} h_i+\frac{1}{2} \Delta t \rho C_p T \mathbf{u}.
\end{equation}

For phase field, the LB evolution equation can be written as
\begin{equation}
	g_{i,p}\left(\mathbf{x}+\mathbf{c}_i \Delta t, t+\Delta t\right)-g_{i,p}(\mathbf{x}, t)=-\frac{1}{\tau_g}\left[g_{i,p}(\mathbf{x}, t)-g_{i,p}^{\mathrm{eq}}(\mathbf{x}, t)\right]+\left(1-\frac{1}{2 \tau_g}\right) \Delta t G_{i,p}(\mathbf{x}, t),
\end{equation}
where $g_{i,p}(\mathbf{x}, t)$ is the distribution function of phase-field variable $\phi_p$, $\tau_g$ is the relaxation parameter related to $ M=c_s^2\left(\tau_g-0.5\right) \delta t $. To recover the conservative AC equation (\ref{eq_ph}), the equilibrium distribution function $g_{i,p}^{\mathrm{eq}}$ and source term $G_{i,p}(\mathbf{x}, t)$ can be designed as
\begin{equation}
	g_{i, p}^{e q}=\omega_{i, p} \phi_p\left(1+\frac{\mathbf{c}_i \cdot \mathbf{u}}{c_{s, p}^2}\right), \quad G_{i, p}=\omega_{i, p}\left[\frac{\mathbf{c}_i \cdot \partial_t\left(\phi_p \mathbf{u}\right)}{c_{s, p}^2}+\mathbf{c}_i \cdot \mathbf{R}_p\right]+\omega_{i, p} \phi_p \nabla \cdot \mathbf{u} ,
\end{equation}
The order parameter is calculated by
\begin{equation}
	\phi_p=\sum_i g_{i,p}+\frac{\Delta t}{2} \phi_p \nabla \cdot \mathbf{u}, \quad \text { for } 1 \leq p \leq N.
\end{equation}

\section{Numerical results and discussion}
\label{sec4}
In this section, some numerical experiments are conducted to demonstrate the capability and reliability of the proposed diffuse-interface method. To this end, the freezing processes of liquid film and pure droplet on the cold substrates are first investigated, and some comparisons between the present work and the previous numerical results, analytical solutions, and experimental data are carried out. Then two challenging problems, including the freezing of compound droplets on supercooled substrate and freezing problems involving impurities, are considered to show the capability of present method in the study of multiphase flows with liquid-solid phase change.

\subsection{Freezing of the liquid film}
\begin{figure}[H]
	\centering
	\includegraphics[scale=0.45]{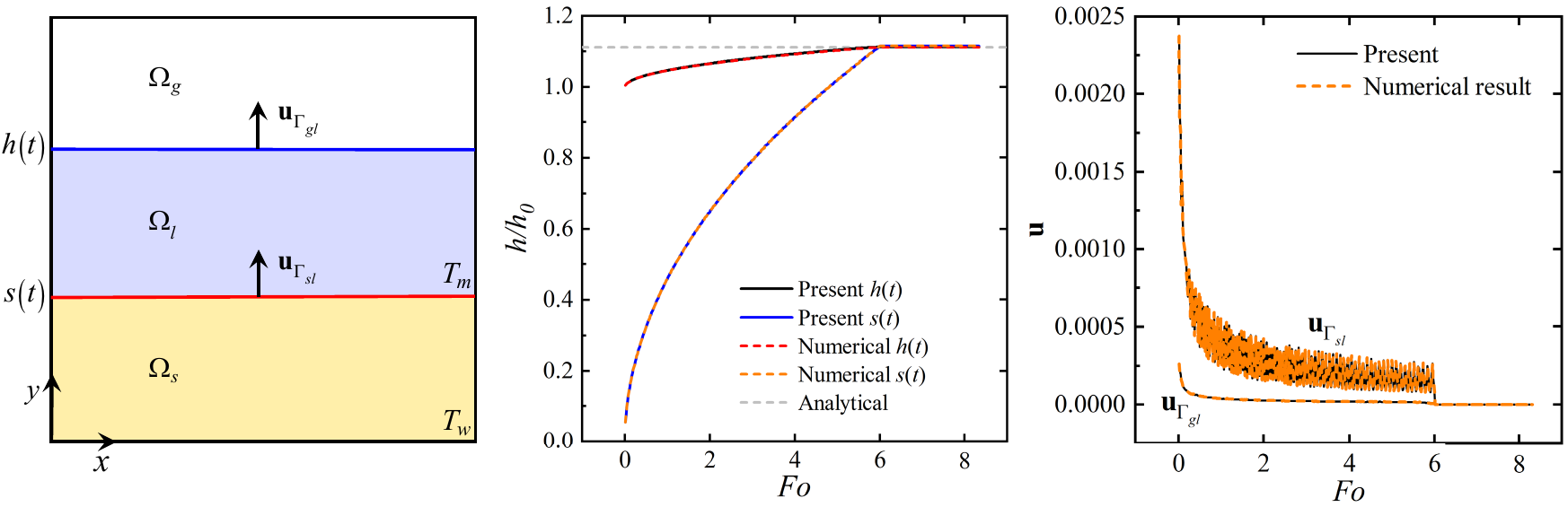}
	\put(-378 ,110){(\textit{a})}
	\put(-258 ,110){(\textit{b})}
	\put(-135 ,110){(\textit{c})}	
	\caption{ (a) Schematic diagram of the single-component liquid film freezing process. The phase-change fluid $\Omega_l$, the ambient fluid $\Omega_g$, and the solid phase $\Omega_s$ formed by freezing are represented by blue, white, and yellow regions. The freezing front $\Gamma_{sl}$ is indicated by red solid line, while the interface $\Gamma_{gl}$ between the phase-change fluid and the ambient fluid is labeled by blue solid line. The positions of these interfaces are denoted by $s(t)$ and $h(t)$, respectively. (b) Comparisons of solid-liquid and gas-liquid interface positions among the present method, reference numerical results \cite{HuangPRE2024,HuangIJHMT2025} and analytical solutions  \cite{LyuJCP2021}. (c) Comparisons of the interfacial velocities predicted by the present method and reported in the previous work \cite{HuangPRE2024}. }
	\label{fig2}
\end{figure}

To test the capability of developed diffuse-interface method  in predicting solidification-induced volume change,  the freezing of a liquid film in the presence of the gas phase is first investigated. As shown in Fig. \ref{fig2}(a), the entire domain is initially maintained at temperature $T_0$, and the liquid film with the height $h_0$ occupies the bottom region, while the remaining domain is filled with the gas phase. Then a low temperature $T_w$ is imposed on the bottom wall, which induces the freezing of the liquid phase at the bottom wall $(y=0)$, and the solid-liquid phase interface $\Gamma_{sl}$ moves toward the liquid phase. In addition, the density difference between the solid and liquid phases leads to a volume change, which in turn leads to the motion of the interface $\Gamma_{gl}$ between the gas phase and the solid–liquid mixture. Once the liquid phase is completely solidified, the height of the solidified liquid is denoted as $h_f$. According to mass conservation of the solid and liquid phases during the freezing process, the final height of the solidified liquid $h_f$ can be obtained from $h_f = \rho_s h_0 / \rho_l$, where $\rho_s$ and $\rho_l$ are the densities of the solid and liquid phases, respectively. For this problem, the Dirichlet boundary conditions $T(x,\ 0,\ t) = T_w$ and $T(x,\infty,\ t) = T_0$ are imposed on the bottom and top boundaries, while the periodic boundary conditions $T(x,\ 0,\ t) = T(x,\ l_y,\ t)$ are applied on the left and right boundaries. Under these conditions, the evolution of the solid phase height satisfies $s(t)=2 \kappa \sqrt{\alpha_s t}$ with $\kappa$ representing the root of the following transcendental equation \cite{LyuJCP2021},
\begin{equation}
	\frac{Ste}{\exp \left(\kappa^2\right) \operatorname{erf}(\kappa)}-\frac{\lambda_l\left(T_0-T_m\right) \sqrt{\alpha_r} Ste}{\lambda_s\left(T_m-T_w\right) \exp \left(\sqrt{\alpha_r} \kappa \rho_r\right)^2 \operatorname{erfc}\left(\sqrt{\alpha_r} \kappa \rho_r\right)}=\kappa \sqrt{\pi},
\end{equation}
where $ \rho_r=\rho_s / \rho_l$ and $\alpha_r=\alpha_s / \alpha_l$ denote the ratios of the density and thermal diffusivity between the solid and liquid phases, respectively. $Ste=C_{p, s}(T_m-T_w)/L$ is the Stefan number. In our simulations, the initial profile of the order parameter is given as
\begin{equation}
	\begin{aligned}
		\phi(x, y)=0.5+0.5 \tanh \frac{\left(h_0-y\right)}{\epsilon / 2},
	\end{aligned}
\end{equation}
where $\epsilon$ is a parameter related to interface thickness. The non-equilibrium extrapolation scheme is used to treat the Dirichlet boundary conditions of temperature field, and the physical parameters are set as $T_b=-1$, $T_m=0$, $T_0=1$, $C_{p,s}/C_{p,l}=1$, $\lambda_{s}/\lambda_{l}=1$. 

We perform some simulations, and show the evolutions of the freezing front and the gas-liquid interface in Fig. \ref{fig2}(b)  where $\rho_s/ \rho_l=0.9$ and $Ste=0.1$, the Fourier number $Fo=\alpha t /l^2$ represents dimensionless time with $l$ being the reference length. From this figure, one can see that the present results are in good agreement with analytical solutions and the previous work \cite{HuangPRE2024}. It is also found that the solid–liquid interface continuously advances upward as freezing progresses, and simultaneously, the gas-liquid interface also moves upward due to volume expansion caused by the density difference between the solid and liquid phases. Furthermore, we also conduct some comparisons of the interfacial velocities [see Fig. \ref{fig2}(c)], and find that the results predicted by the present method agree well with those reported in some previous studies \cite{HuangPRE2024,HuangIJHMT2025}. To see the differences between the present results and analytical solutions, we present a quantitative comparison in Table \ref{table1}, where the maximum error is less than 5$\%$, which demonstrates that the proposed diffuse-interface model can accurately capture the volume expansion associated with phase change, and strictly satisfy the principle of mass conservation. 

\begin{table}[H]
	\centering
	\caption{ Comparisons of the dimensionless solid-phase height $h_f/h_0$ between numerical results and analytical solutions at different values of $\rho_s / \rho_l$. }
	\begin{tabular}{ccccclccc}
		\hline \hline
		\multirow{2}{*}{$\rho_l / \rho_s$} & \multirow{2}{*}{Analytical} & \multicolumn{3}{c}{Numerical}      &  & \multicolumn{3}{c}{Relative error}                                    \\ \cline{3-5} \cline{7-9} 
		&                             & $Ste=0.1$ & $Ste=0.15$ & $Ste=0.2$ &  & $Ste=0.1$             & $Ste=0.15$            & $Ste=0.2$             \\ \cline{1-5} \cline{7-9} 
		0.7                            & 1.4286                      & 1.4537    & 1.4594     & 1.4634    &  & $1.76 \times 10^{-2}$ & $2.15 \times 10^{-2}$ & $2.44 \times 10^{-2}$ \\
		0.75                           & 1.3333                      & 1.3582    & 1.3586     & 1.3589    &  & $1.87 \times 10^{-2}$ & $1.90 \times 10^{-2}$ & $1.92 \times 10^{-2}$ \\
		0.8                            & 1.2500                      & 1.2612    & 1.2612     & 1.2612    &  & $8.93 \times 10^{-3}$ & $8.92 \times 10^{-3}$ & $8.94 \times 10^{-3}$ \\
		0.85                           & 1.1765                      & 1.1806    & 1.1806     & 1.1806    &  & $3.53 \times 10^{-3}$ & $3.53 \times 10^{-3}$ & $3.53 \times 10^{-3}$ \\
		0.9                            & 1.1111                      & 1.1121    & 1.1121     & 1.1121    &  & $8.90 \times 10^{-4}$ & $8.81 \times 10^{-4}$ & $8.81 \times 10^{-3}$ \\
		0.95                           & 1.0526                      & 1.0526    & 1.0526     & 1.0526    &  & $5.85 \times 10^{-5}$ & $5.85 \times 10^{-5}$ & $5.85 \times 10^{-5}$ \\
		1                              & 1.0000                      & 1.0000    & 1.0000     & 1.0000    &  & $4.90 \times 10^{-7}$ & $3.10 \times 10^{-7}$ & $1.57 \times 10^{-6}$ \\ \hline \hline
	\end{tabular}
	\label{table1}
\end{table}

\begin{figure}[H]
	\centering
	\includegraphics[scale=0.4]{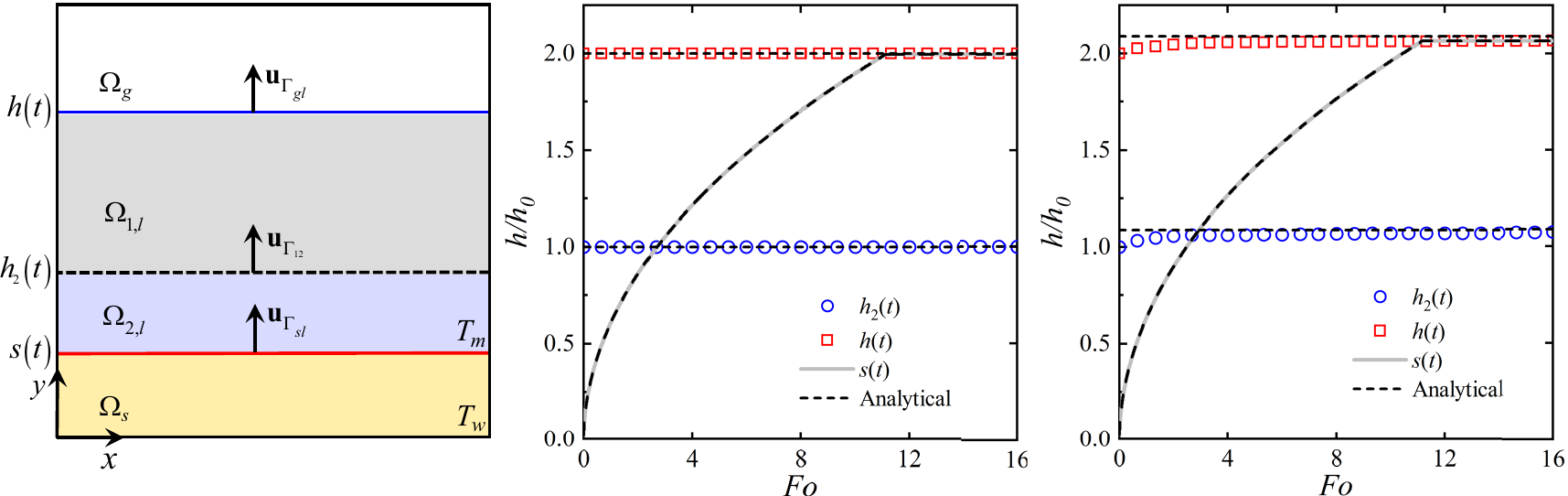}
	\put(-331 ,100){(\textit{a})}
	\put(-220 ,100){(\textit{b})}
	\put(-107 ,100){(\textit{c})}	
	\caption{ (a) Schematic of freezing of a liquid pool composed of immiscible liquid fluids. Fluid 1 $(\Omega_{1,l})$, fluid 2 $(\Omega_{2,l})$, and the ambient fluid $(\Omega_{g})$ are represented by gray, blue, and white, respectively, while the solid phase $(\Omega_{s})$ formed by freezing is indicated in yellow. The functions $s(t)$, $h_2(t)$, and $h(t)$ denote the positions of the freezing front, the interface between fluid 1 and fluid 2, and the interface between the ambient fluid and fluid 1, respectively. Temporal evolutions of the solidification front $\Gamma_{sl}$, fluid-fluid interface $\Gamma_{12}$, and gas-liquid interface $\Gamma_{gl}$ under different density ratios (b) $\rho_1^s/ \rho_1^l=1$, $\rho_2^s/ \rho_2^l=1.0$ and (c) $\rho_1^s/ \rho_1^l=1$ and $\rho_2^s/ \rho_2^l=0.95$. }
	\label{fig3}
\end{figure}

To further test the diffuse-interface method for solid-liquid phase change in the multiphase system, we consider the problem in Fig. \ref{fig3}(a). Initially, a liquid column composed of two immiscible components is full of the bottom region, while the upper region is occupied by the gas phase. The initial distributions of order parameters are given by
\begin{equation}
	\begin{aligned}
		\phi_2(x, y) &= \frac{1}{2} + \frac{1}{2} \tanh \frac{\left(h_0 - y\right)}{\epsilon / 2} , \\
		\phi_1(x, y) &= [1 - \phi_2(x, y)] \left[ \frac{1}{2} + \frac{1}{2} \tanh \frac{\left(2h_0 - y\right)}{\epsilon / 2}  \right], \\
		\phi_3(x, y) &= 1 - \phi_1(x, y) - \phi_2(x, y),
	\end{aligned}
\end{equation}
where $h_0$ is the initial height of each fluid layer. Apart from the initial configuration, the simulation settings are consistent with those of the previous case. Fig. \ref{fig3}(b) shows the temporal evolutions of the freezing front $\Gamma_{sl}$, the fluid-fluid interface $\Gamma_{12}$, and the gas-liquid interface $\Gamma_{gl}$ for the case of $\rho_1^s/ \rho_1^l=1$ and $\rho_2^s/ \rho_2^l=1$. From this figure, one can observe that the freezing front $\Gamma_{sl}$ advances continuously upward, and the numerical results are close to the analytical solutions. Since the densities of both fluids 1 and 2 are identical to each other, no volume change occurs, and thus the positions of the interfaces $\Gamma_{12}$ and $\Gamma_{gl}$ remain fixed throughout the freezing process. Fig. \ref{fig3}(c) presents the results of the case $\rho_1^s/ \rho_1^l=1$ and $\rho_2^s/ \rho_2^l=0.95$. In this case, the density difference of fluid 2 before and after solidification induces a volume expansion, which causes the interface $\Gamma_{12}$ to move upward. This displacement, in turn, drives the movement of the gas–liquid interface $\Gamma_{gl}$. In contrast, since the freezing of fluid 1 undergoes no volume change, its height remains constant during the freezing process.
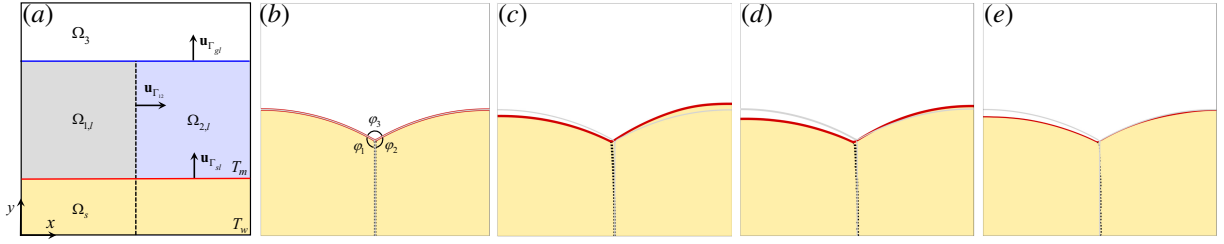
\begin{figure}[H]
	\centering
	\includegraphics[scale=0.3]{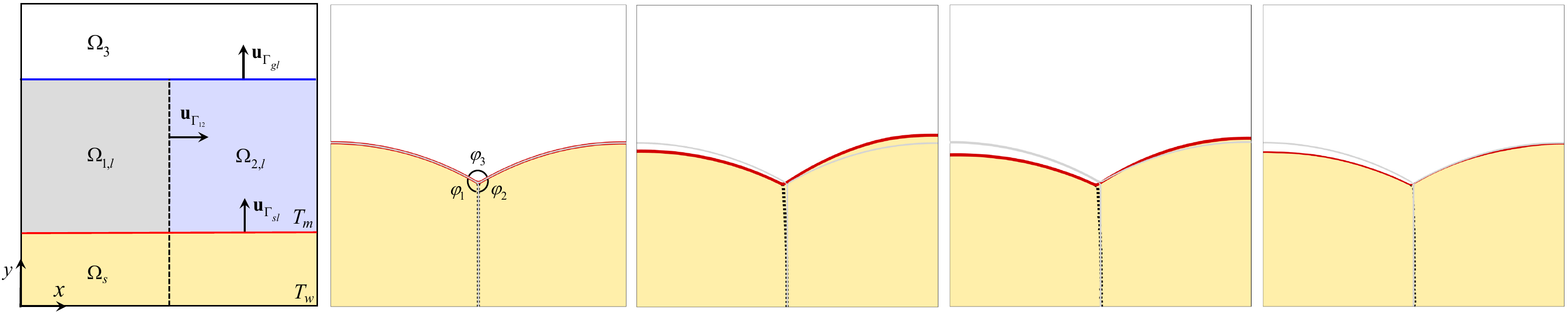}
	\put(-448 ,82){(\textit{a})}
	\put(-359.5 ,82){(\textit{b})}
	\put(-270 ,82){(\textit{c})}	
	\put(-179 ,82){(\textit{d})}
	\put(-88 ,82){(\textit{e})}	
	\caption{ (a) Schematic of a liquid pool freezing process. The liquid pool is composed of two immiscible liquid fluids ($\Omega_{1,l}$, $\Omega_{2,l}$) distributed on the left and right sides, where fluid 1 is marked in gray and fluid 2 in blue. The solid phase $\Omega_{s}$ formed by freezing is denoted by yellow. Comparisons of phase interfaces before and after freezing for (b) $\rho_1^s/ \rho_1^l=1$, $\rho_2^s/ \rho_2^l=1$, (c) $\rho_1^s/ \rho_1^l=1.05$, $\rho_2^s/ \rho_2^l=0.95$, (d) $\rho_1^s/ \rho_1^l=1.1$, $\rho_2^s/ \rho_2^l=1.05$ and (e) $\rho_1^s/ \rho_1^l=1.1$, $\rho_2^s/ \rho_2^l=1$, where the gray line is the initial phase interface contour, and the red line and the black dashed line are the freezing front and the two phase interfaces after freezing. }
	\label{fig4}
\end{figure}

Next, we examine the freezing process of a liquid pool composed of two laterally distributed fluids, as shown in Fig. \ref{fig4}(a), the initial distributions of the order parameter are given by
\begin{equation}
	\begin{aligned}
		\phi_1(x, y) &= \left[ \frac{1}{2} + \frac{1}{2} \tanh  \frac{\left(L/2 - x\right)}{\epsilon / 2}  \right] 
		\times \left[ \frac{1}{2} + \frac{1}{2} \tanh  \frac{\left(L/2 - y\right)}{\epsilon / 2}  \right], \\
		\phi_2(x, y) &= \left[ \frac{1}{2} + \frac{1}{2} \tanh  \frac{\left(x - L/2\right)}{\epsilon / 2}  \right] 
		\times \left[ \frac{1}{2} + \frac{1}{2} \tanh  \frac{\left(L/2 - y\right)}{\epsilon / 2}  \right], \\
		\phi_3(x, y) &= 1 - \phi_1(x, y) - \phi_2(x, y),
	\end{aligned}
\end{equation}
where $L$ is the size of the square computational domain. In our simulations, the surface tension coefficients are set as $\sigma_{23}=\sigma_{31}=\sigma_{12}=0.01$. Additionally, under the balance condition (\ref{eq0}), the final equilibrium interfacial angles can be determined as $\varphi_1=\varphi_2=\varphi_3=120^\circ$. We perform some simulations, and present a comparison between the initial distribution and final solidified contours of the liquid pool under different solid-to-liquid density ratios in Figs. \ref{fig4}(b)-(e). As shown in Fig. \ref{fig4}(b) where $\rho_1^s/\rho_1^l=1$ and $\rho_2^s/\rho_2^l=1$, the volume remains unchanged during solidification process, and the numerical results of the phase distribution are in excellent agreement with the theoretical predictions. For the case of $\rho_1^s/ \rho_1^l=1.05$ and $\rho_2^s/ \rho_2^l=0.95$ [see Fig. \ref{fig4}(c)], the fluid 1 undergoes the volume contraction, while the fluid 2 experiences the volume expansion. For the case of $\rho_1^s/ \rho_1^l=1.1$ and $\rho_2^s/ \rho_2^l=1.05$ [see Fig. \ref{fig4}(d)], both the fluid 1 and the fluid 2 undergo the volume contraction. When $\rho_1^s/ \rho_1^l=1.1$ and $\rho_2^s/ \rho_2^l=1.0$ [see Fig. \ref{fig4}(e)], the fluid 1 undergoes the volume contraction, while the fluid 2 maintains a constant volume. In summary, the numerical results demonstrate that the volume change is strictly governed by the density ratio, i.e., the solidification results in the volume expansion when $\rho_s/\rho_l < 1$, volume contraction when $\rho_s/\rho_l > 1$, and no net volume change when $\rho_s/\rho_l = 1$.

\begin{figure}[H]
	\centering
	\includegraphics[scale=0.29]{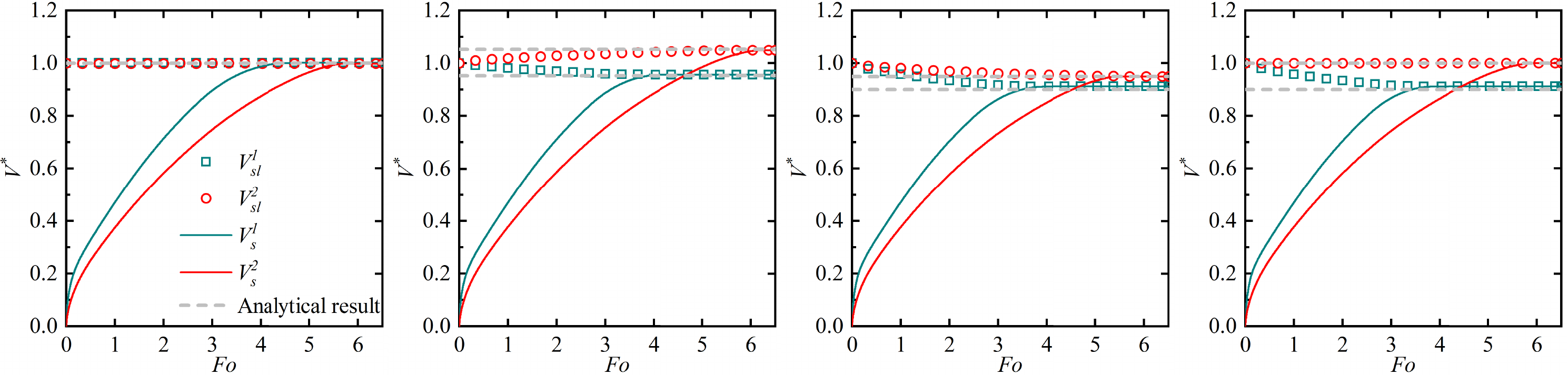}
	\put(-419, 93){(\textit{a})}
	\put(-314.5, 93){(\textit{b})}
	\put(-209.5, 93){(\textit{c})}	
	\put(-106, 93){(\textit{d})}
	\caption{ The evolutions of normalized phase volumes during the freezing processes under different density ratios (a) $\rho_1^s/ \rho_1^l=1$, $\rho_2^s/ \rho_2^l=1$, (b) $\rho_1^s/ \rho_1^l=1.05$, $\rho_2^s/ \rho_2^l=0.95$, (c) $\rho_1^s/ \rho_1^l=1.1$, $\rho_2^s/ \rho_2^l=1.05$ and (d) $\rho_1^s/ \rho_1^l=1.1$, $\rho_2^s/ \rho_2^l=1$. $V_s^p$ is the volume of the solid phase formed by the freezing of fluid $p$, and $V_{sl}^p$ is the total volume of the solid and liquid phases of fluid $p$, which are both normalized by the initial volume. }
	\label{fig5}
\end{figure}
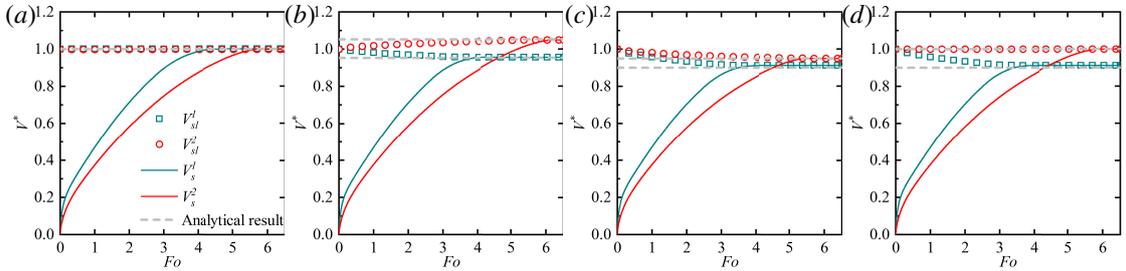

In addition, we also conduct some quantitative comparisons of the volume evolutions of both the solid phase and the total solid-liquid mixture of fluids 1 and 2 in Fig. \ref{fig5} where four different cases are considered. It is found from this figure that the results are consistent with above analysis, and are also in excellent agreement with the theoretical predictions. These results demonstrate that the present diffuse-interface method can accurately capture the solidification process in multiphase fluids with distinct volume changes of different phases.

\subsection{Freezing of the pure sessile droplet}

\begin{figure}[H]
	\centering
	\includegraphics[scale=0.45]{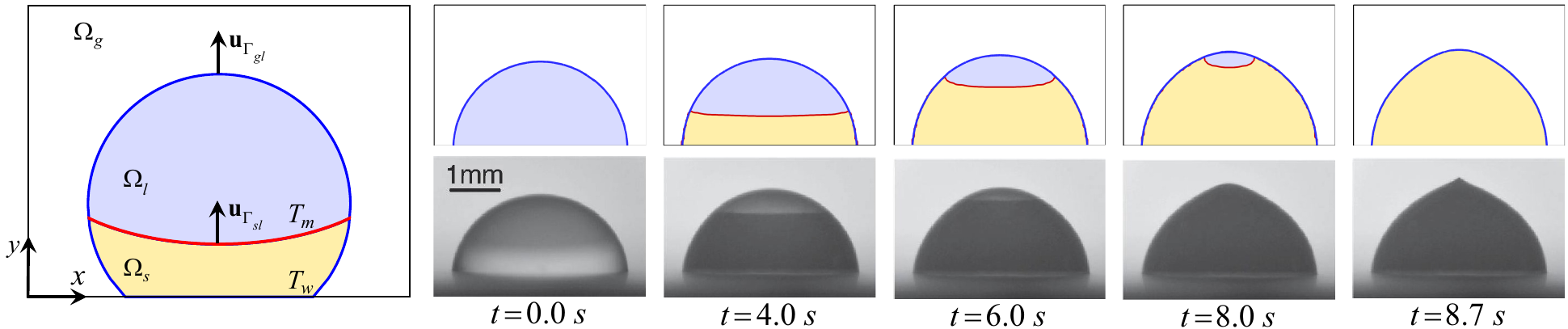}
	\put(-425, 80){(\textit{a})}
	\put(-303, 80){(\textit{b})}
	\caption{ (a) Schematic of a sessile droplet freezing on a cold substrate, the blue region represents the liquid phase $\Omega_{l}$, the white region indicates the surrounding gas phase $\Omega_{g}$, and the yellow region denotes the solid phase $\Omega_{s}$ formed during the solidification process, the red line denotes the freezing front $\Gamma_{sl}$, and the blue solid line represents the interface $\Gamma_{gl}$ between the solid-liquid mixture and the surrounding gas phase. (b) Some comparisons  between experimental data and numerical results of the freezing droplet at different times. }
	\label{fig6}
\end{figure}

In this part, we investigate the freezing of a single water droplet on a supercooled substrate, as illustrated in Fig. \ref{fig6}(a). Initially, the droplet is placed on the substrate, and the temperature of the entire domain is maintained at a uniform temperature $T_0$. After the droplet reaches the specified equilibrium contact angle $\theta$, a constant low temperature $T_w$ is imposed on the bottom substrate, triggering the solidification from the droplet base. In our simulations, the periodic boundary conditions are applied along the horizontal direction. Additionally, for the flow field, the no-slip boundary condition is applied on both the top and bottom walls, while for the temperature field, the top boundary is adiabatic, the bottom substrate is maintained at a constant temperature $T_w$. The initial distribution of the phase-field order parameter is given by 
\begin{equation}
	\phi_1(x, y)=\frac{1}{2}+\frac{1}{2} \tanh \frac{R-\sqrt{(x-x_0)^2+(y-y_0)^2}}{\epsilon / 2},
\end{equation}
where $(x_0,y_0)$ and $R$ are the center position and radius of the droplet. Some other simulation parameters are set to be $\rho_s / \rho_l=0.92$, $C_{p, s}/C_{p, l}=1.0$, $Ste=0.15$, $\theta=87^{\circ}$.

Fig. \ref{fig6}(b) illustrates the evolutions of the freezing front and droplet profile during the freezing process. As seen from this figure, the present results agree well with the experimental data \cite{ZhangETFS2017} and successfully capture the volume expansion caused by the solid-liquid density difference, which leads to the formation of a conical tip at the droplet apex. However, a notable discrepancy is also observed in the final droplet morphology: the experimental result shows a sharp tip formed at the top of droplet, whereas the numerical results exhibit a smooth and rounded cap. We would like to point out that this outcome is consistent with the previous numerical results of Lyu et al. \cite{LyuJCP2021} and Wei et al. \cite{WeiJCP2025}, where the rounded cap morphology is also reporeted. We also note that the theoretical analysis by Snoeijer et al. \cite{SnoeijerAJP2012} demonstrated that a tip singularity (i.e., the sharp tip) only occurs when the solid-to-liquid density ratio satisfies the condition $\rho_s/\rho_l<0.75$, which does not hold in the present study. On the other hand, the difference in tip sharpness between experimental and numerical results may also be attributed to the presence of dissolved gas in the droplet. Actually, Chu et al. \cite{ChuPRF2019} experimentally revealed the formation of numerous micro-bubbles during the freezing of a pure water droplet, which results in a porous ice structure, effectively reducing the actual effective density of ice. As a consequence, the effective density of ice may fall below the theoretical threshold of $\rho_s/\rho_l=0.75$, thus enabling the formation of sharp tip, as observed in experiments. This hypothesis is further supported by a recent numerical work \cite{WeiJCP2025}, which demonstrated that the sharp-tip morphology could be reproduced when bubble-induced reduction in effective density is taken into account. In contrast, the numerical results with a fixed density ratio of $\rho_s/\rho_l=0.92$ produce the smooth and rounded caps which are in agreement with the present results. To confirm this statement, we present some comparisons of the evolutions of the freezing front and droplet height, as well as the final solidified contours, with both experimental data and reference numerical results, and find that the results are in excellent agreement with both experimental and numerical results \cite{WeiJCP2025,ZhangETFS2017}.
\begin{figure}[H]
	\centering
	\includegraphics[scale=0.5]{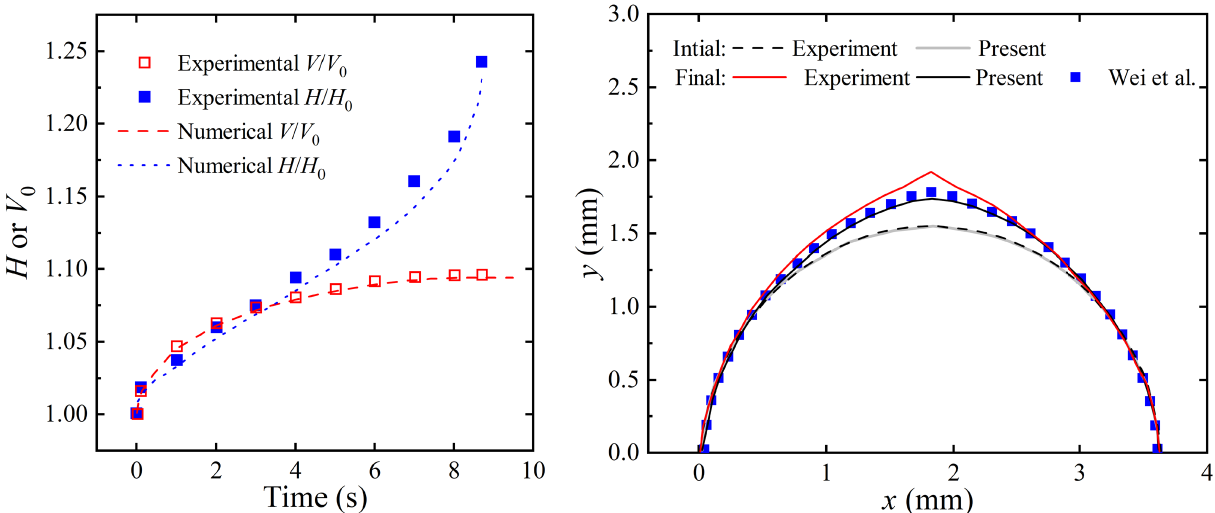}
	\put(-295, 115){(\textit{a})}
	\put(-160, 115){(\textit{b})}
	\caption{ (a) Comparisons of the temporal evolutions of normalized droplet height $(H/H_0)$ and volume $(V/V_0)$ between the experimental data and the present numerical results. (b) A comparison of droplet profiles before and after freezing among the experiment data \cite{ZhangETFS2017}, numerical results \cite{WeiJCP2025}, and the present work. }
	\label{fig7}
\end{figure}

\begin{figure}[H]
	\centering
	\includegraphics[scale=0.4]{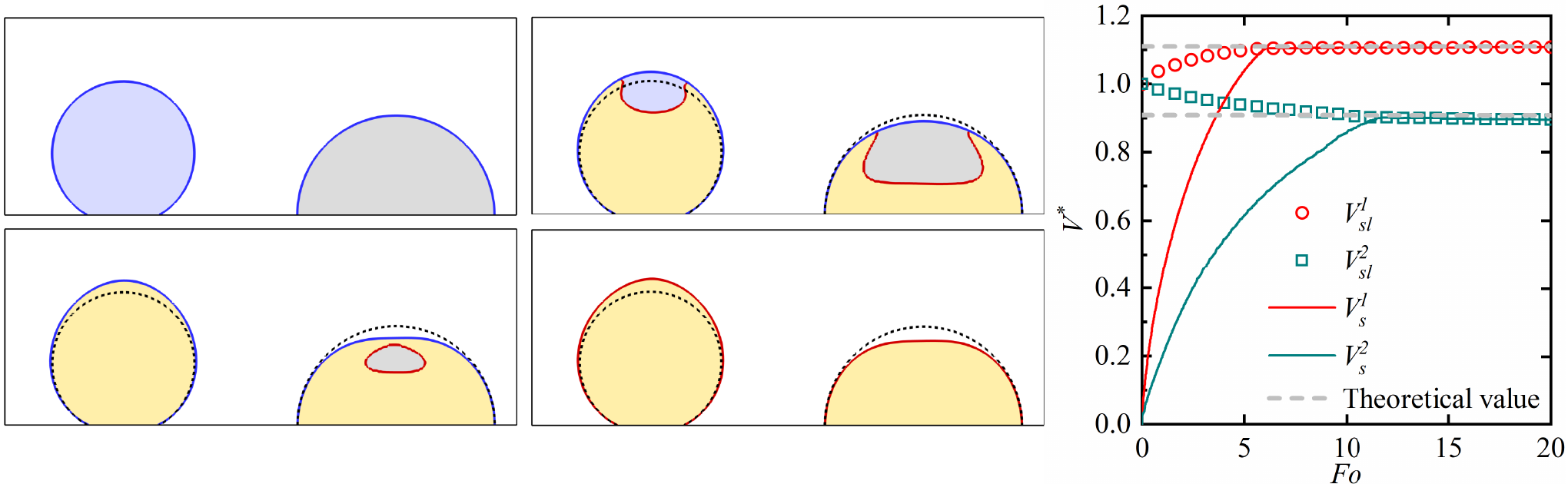}
	\put(-405, 109){(\textit{a})}
	\put(-130, 109){(\textit{b})}
	\put(-392, 109){$Fo=0.0$}
	\put(-260, 109){$Fo=2.7$}	
	\put(-392, 56){$Fo=8.0$}
	\put(-260, 56){$Fo=11.7$}		
	\caption{ (a) The freezing process of two droplets on a cold substrate, where blue, gray, white, and yellow regions represent the droplet 1, droplet 2, environmental fluid 3, and the solid phase formed by freezing, respectively. The blue line denotes the interface between the droplets and the surrounding environmental fluid, and the red line indicates the freezing front. (b) Evolution of the volume of each phase over time during the freezing process. }
	\label{fig8}
\end{figure}
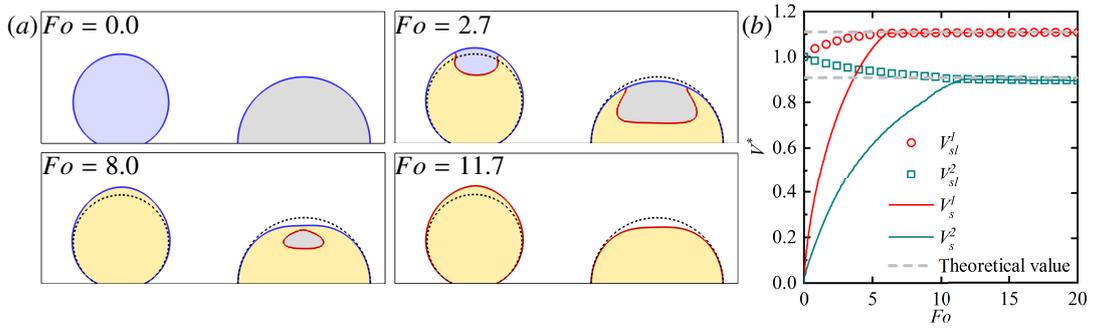

To further evaluate the capability of the diffuse-interface method in the study of the freezing behavior of multiple droplets, we investigate the freezing process of two distinct droplets located on a cold substrate. As shown in Fig. \ref{fig8}(a) at $Fo=0.0$, the droplets 1 and 2 are initially placed on a cold substrate with two specified contact angles $\theta_1=150^{\circ}$ and $\theta_2=90^{\circ}$, respectively. The boundary conditions remain consistent with the previous case, and other key simulation parameters are set as $Ste=0.2$, $\rho_1^s/\rho_1^l=0.9$, $\rho_1^s/\rho_1^l=1.1$, $\sigma_{pq}=0.01$ ($1 \leq p \neq q \leq 3$), $\alpha_3/\alpha_1=3$, $\alpha_3/\alpha_2=1.5$. As shown in Fig. \ref{fig8}(a), the droplet 1 with $\rho_1^s/\rho_1^l<1.0$ undergoes the volume expansion during the solidification process, while the droplet 2 with $\rho_1^s/\rho_1^l>1.0$ experiences the volume contraction. During the freezing process, the solidification rate of the triple line increases due to the high thermal diffusivity of the surrounding fluid 3, resulting in an upward-curving solid–liquid interface near the edges, while the central interface remains relatively flat. This behavior, commonly referred to as secondary solidification, has been previously reported in the previous works \cite{MohammadipourJFM2024, LyuDroplet2023}. For this reason, the solid shell is formed for the droplet 2, and encapsulates the internal liquid phase. Then the volume change of the encapsulated internal liquid can lead to two possible outcomes. The first is that when the solid shell is sufficiently strong, internal expansion ($\rho_1^s/\rho_1^l<1.0$) may generate residual stress within the droplet, while the contraction ($\rho_1^s/\rho_1^l>1.0$) may result in the formation of internal pores. The second is that when the expansion pressure due to $\rho_1^s/\rho_1^l<1.0$ exceeds the strength of the outer solid layer, cracks may form in the shell, allowing liquid to escape and solidify on the surface. Although these phenomena have been reported in the previous studies \cite{MohammadipourJFM2024, LyuDroplet2023}, the present diffuse-interface method (assuming the solid phase is rigid and non-breakable) cannot reproduce the complex processes involving deformation or failure of the solid phase, such as shell rupture, liquid exudation, or pore formation, as described above. Although there is this limitation, the method can accurately reproduce the overall morphological evolution, and the characteristics are similar to those observed for a single-droplet freezing \cite{GuoIJTS2024}. Besides, we also give a comparison of the volume changes of the two droplets throughout the freezing process, and find that the present results are in good agreement with theoretical predictions and satisfy the mass conservation. These findings indeed demonstrate that the present diffuse-interface method effectively captures the key freezing characteristics of droplets with different solid-to-liquid density ratios.

\subsection{Freezing of the compound sessile droplet}

The freezing of compound sessile droplet, composed of two or more immiscible fluids adhering to a cold substrate, are commonly encountered in various industrial applications \cite{ZhanCICP2024,ZhangJFM2021}. These droplets can exhibit a range of complex morphological configurations depending on the constituent components and wettability of the substrate. Usually, they can be classified into several typical types when the droplet are in contact with the substrate, including collar, encapsulated, Janus, and lens modes \cite{ZhangJFM2021}, which are provided in Fig. \ref{fig10} where $Fo=0.0$. In the collar configuration, the droplet 1 (composed of fluid 1) is partially immersed in the droplet 2 (composed of fluid 2), while in the encapsulated configuration, the droplet 1 is fully enveloped by the droplet 2 in an axisymmetric manner, and both droplets are in contact with the substrate. In the Janus configuration, the droplet 1 sits on the top of droplet 2, with both droplets in contact with the substrate. In contrast, in the lens configuration, the droplet 1 floats on the top of droplet 2, but only the droplet 2 is in contact with the substrate. In the following, we will investigate the freezing dynamics of such compound sessile droplets on a cold substrate, focusing on how interfacial interactions and solid-liquid density ratio influence the final solidified morphology.

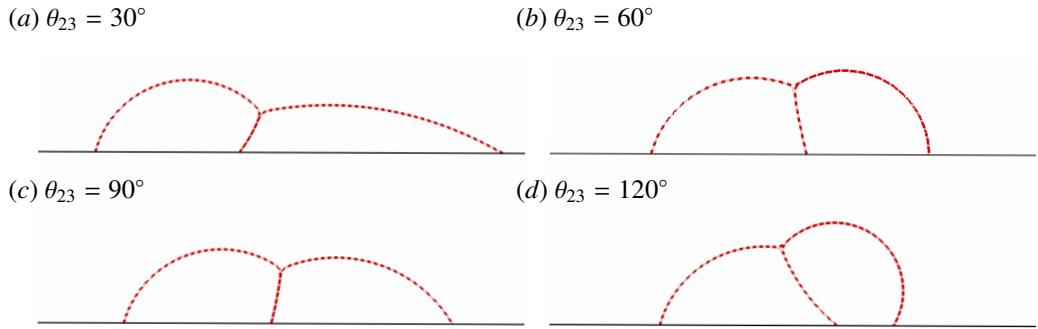
\begin{figure}[H]
	\centering
	\includegraphics[width=0.8\textwidth]{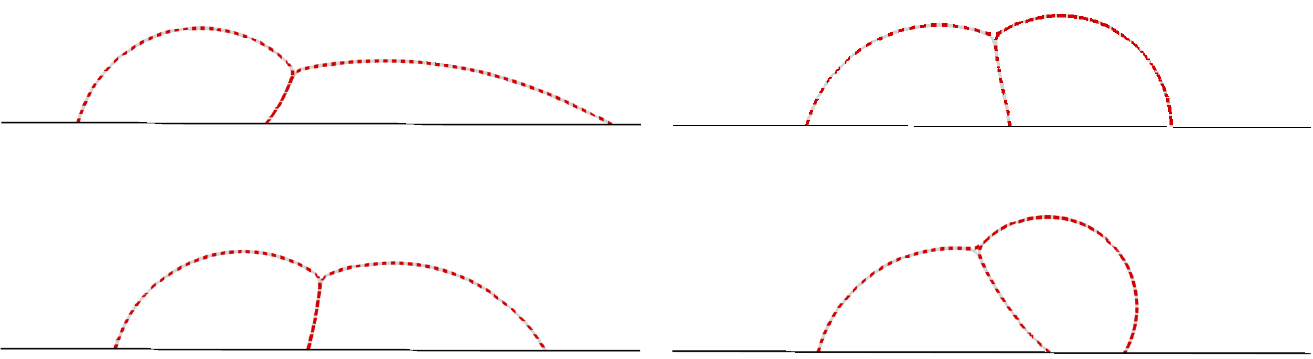}
	\put(-385, 115){(\textit{a}) $\theta_{23}=30^{\circ}$}
	\put(-195, 115){(\textit{b}) $\theta_{23}=60^{\circ}$}	
	\put(-385, 50){(\textit{c}) $\theta_{23}=90^{\circ}$}
	\put(-195, 50){(\textit{d}) $\theta_{23}=120^{\circ}$}		
	
	\caption{ Comparisons of present results (red dashed) and previous work \cite{ZhanCICP2024} (gray solid) for the equilibrium distributions of a Janus compound droplet on a substrate with different values of $\theta_{23}$ ($\theta_{13}=75^{\circ}$). }
	\label{fig9}
\end{figure}

We first test the wetting boundary conditions for \textit{N}-phase flows by simulating the spreading of a Janus compound droplet on a flat substrate, as illustrated in Fig. \ref{fig1}(a). In the computational domain $[-150,150] \times[0,150]$, a semicircular compound droplet with the radius $R = 60$ is initially placed on the bottom wall, and the order parameters are initialized by \cite{ZhanCICP2024}
\begin{equation}
	\begin{aligned}
		\phi_1(x,y) &= \left( \frac{1}{2} + \frac{1}{2} \tanh \frac{R - \sqrt{x^2 + y^2}}{\epsilon / 2} \right) \left( \frac{1}{2} - \frac{1}{2} \tanh \frac{2x}{\epsilon} \right), \\
		\phi_2(x,y) &= \left( \frac{1}{2} + \frac{1}{2} \tanh \frac{R - \sqrt{x^2 + y^2}}{\epsilon / 2} \right) \left( \frac{1}{2} + \frac{1}{2} \tanh \frac{2x}{\epsilon} \right). 
	\end{aligned}
\end{equation}
The physical boundary conditions are the same as those in the previous section, and additionally, the wetting boundary conditions adopted in this study are the same as those in \cite{ZhanCICP2024}. In numerical simulations, the parameters are set as $\rho_1:\rho_2:\rho_3:\rho_4=1000:500:100:1$, $\mu_1:\mu_2:\mu_3:\mu_4=100:50:10:0.1$, $\epsilon=5\Delta x$, $\sigma_{pq}=0.01$ ($1 \leq p \neq q \leq 3$). We perform several simulations under different values of $\theta_{23}$ ($\theta_{13}=75^{\circ}$), and show the droplet morphologies in Fig. \ref{fig9}. As the contact angle $\theta_{23}$ increases, the wetting length $L_2$ of fluid 2 (right side) decreases significantly, while the wetting length $L_1$ of fluid 1 (left side) increases slightly. To quantitatively evaluate the accuracy of the diffuse-interface method, the numerical wetting lengths of fluids 1 and 2 are measured, and compared with analytical solutions and previous results in Table \ref{table2}. From this table, one can see that the present results are in a good agreement with both analytical predictions and previous numerical data, confirming the accuracy of the implemented wetting boundary conditions.

\begin{table}[H]
	\centering
	\caption{ The equilibrium spreading lengths $L_1$ and $L_2$ (normalized by the initial radius $R$) of a compound droplet with $\theta_{13}=75^{\circ}$. }	
	\begin{tabular}{ccclcclcc}
		\hline \hline
		\multirow{2}{*}{Case}                      & \multirow{2}{*}{Length} & \multirow{2}{*}{Analytical} & \multirow{2}{*}{} & \multicolumn{2}{c}{Zhan et al. \cite{ZhanCICP2024}}                                 & \multirow{2}{*}{} & \multicolumn{2}{c}{Present}                                     \\ \cline{5-6} \cline{8-9} 
		&                         &                             &                   & \multicolumn{1}{l}{Result} & \multicolumn{1}{l}{Relative error} &                   & \multicolumn{1}{l}{Result} & \multicolumn{1}{l}{Relative error} \\ \hline
		$\theta_{23}=30^{\circ}$                   & $L_1$                   & 1.3482                      & \multirow{8}{*}{} & 1.3170                     & 2.31E-02                           & \multirow{8}{*}{} & 1.3273                     & 1.55E-02                           \\
		& $L_2$                   & 2.4862                      &                   & 2.5928                     & 4.29E-02                           &                   & 2.4332                     & 2.13E-02                           \\
		\multirow{2}{*}{$\theta_{23}=60^{\circ}$}  & $L_1$                   & 1.3563                      &                   & 1.3286                     & 2.04E-02                           &                   & 1.3573                     & 7.17E-04                           \\
		& $L_2$                   & 1.6666                      &                   & 1.6512                     & 9.20E-03                           &                   & 1.6693                     & 1.64E-03                           \\
		\multirow{2}{*}{$\theta_{23}=90^{\circ}$}  & $L_1$                   & 1.4328                      &                   & 1.4092                     & 1.65E-02                           &                   & 1.4322                     & 4.52E-04                           \\
		& $L_2$                   & 1.1245                      &                   & 1.1183                     & 5.50E-03                           &                   & 1.1232                     & 1.14E-03                           \\
		\multirow{2}{*}{$\theta_{23}=120^{\circ}$} & $L_1$                   & 1.6213                      &                   & 1.6313                     & 6.20E-03                           &                   & 1.6292                     & 4.87E-03                           \\
		& $L_2$                   & 0.5334                      &                   & 0.4988                     & 6.49E-02                           &                   & 0.5172                     & 3.05E-02                           \\ \hline \hline
	\end{tabular}
	\label{table2}
\end{table}

Next, we study the freezing behavior of the composite sessile droplet described above. The initialization of all order parameters for each case is summarized in Table \ref{Tab33}, and both, fluid 1 and fluid 2 are initialized with identical volumes. Once the composite sessile droplet reaches the specified contact angle (see $Fo = 0.0$ in Fig. \ref{fig10}), a low temperature is imposed on the bottom substrate. In our simulations, the specific parameters are given by $\theta_{13}=90^{\circ}$, $\theta_{23}=90^{\circ}$, $\rho_1^s / \rho_1^l=1.1$, $\rho_2^s / \rho_2^l=0.9$, $\sigma_{pq}=0.01$ $(p \neq q)$, $Ste=0.2$, $\alpha_{2} / \alpha_{1} = 3$, $\alpha_{3}/ \alpha_{1} = 1.5$. 

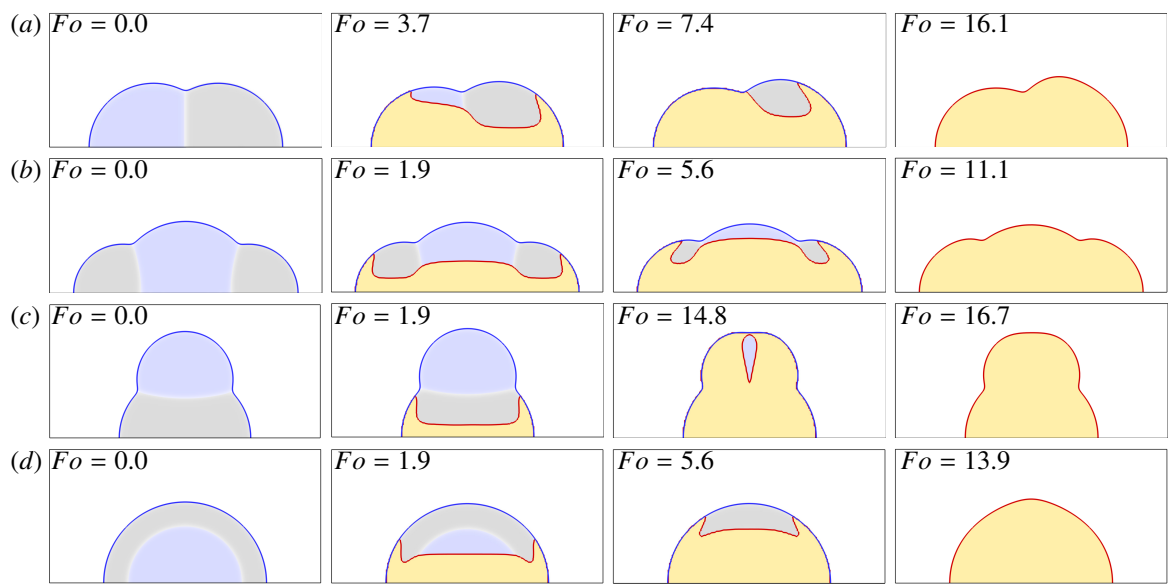
\begin{figure}[H]
	\centering
	\includegraphics[width=0.9\textwidth]{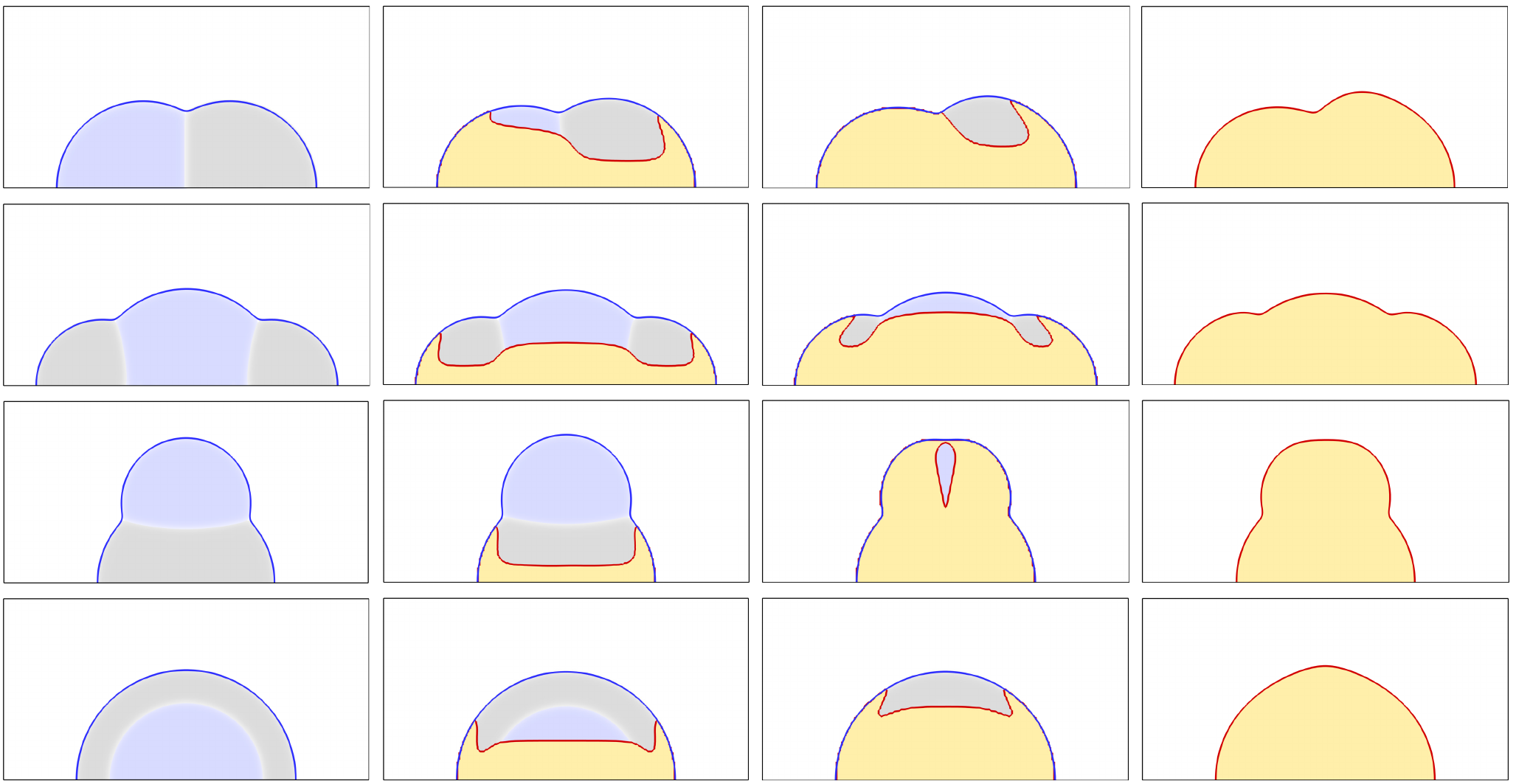}
	\put(-435, 209){(\textit{a})}
	\put(-435, 154){(\textit{b})}	
	\put(-435, 99){(\textit{c})}	
	\put(-435, 44){(\textit{d})}
	\put(-420, 209){$Fo=0.0$}
	\put(-314, 209){$Fo=3.7$}	
	\put(-208, 209){$Fo=7.4$}	
	\put(-102, 209){$Fo=16.1$}	
	\put(-420, 154){$Fo=0.0$}
	\put(-314, 154){$Fo=1.9$}	
	\put(-208, 154){$Fo=5.6$}	
	\put(-102, 154){$Fo=11.1$}		
	\put(-420, 99){$Fo=0.0$}
	\put(-314, 99){$Fo=1.9$}	
	\put(-208, 99){$Fo=14.8$}	
	\put(-102, 99){$Fo=16.7$}			
	\put(-420, 44){$Fo=0.0$}
	\put(-314, 44){$Fo=1.9$}	
	\put(-208, 44){$Fo=5.6$}	
	\put(-102, 44){$Fo=13.9$}			
	\caption{ Freezing dynamics of four different compound droplets on a cryogenic substrate: (a) Janus configuration, (b) collar configuration, (c) lens configuration, (d) encapsulated configuration, where the gray, blue, white, and yellow regions denote droplet 1, droplet 2, ambient fluid and solid phase, respectively. The red solid line represents the freezing front, and the blue solid line denotes the interface between the droplet and the surrounding fluid. }
	\label{fig10}
\end{figure}

Fig. \ref{fig10} illustrates the freezing dynamics of the composite droplet with four different configurations, i.e., Janus, collar, lens, and encapsulated modes. It is found from this figure that during the freezing process, the freezing front (indicated by a red solid line) advances upward and is significantly influenced by the fluid properties. Owing to the higher thermal diffusivity of the ambient fluid, secondary freezing phenomena occur in all configurations, with accelerated solidification at the triple contact point \cite{MohammadipourJFM2024,LyuDroplet2023,HuangIJHMT2025}. Because the fluid 1 (gray) has a higher thermal diffusivity than fluid 2 (blue), it consistently freezes more rapidly for all configurations, although each configuration exhibits different freezing path. In the Janus configuration, the droplet 1 solidifies preferentially, while the solidification of droplet 2 lags and is constrained by the already solidified interface. In the collar configuration, the annular droplet 2 forms an efficient radial heat conduction path due to its lateral contact with the ambient fluid. In the lens configuration, rapid solidification at the triple-phase point leads to the formation of a solid shell around droplet 1, similar to the structure observed at $Fo=8.0$ in Fig. \ref{fig8}. In the encapsulated configuration, the droplet freezing characteristics are similar to those of pure droplets.

\begin{table}[H]
	\centering
	\caption{ Initial distributions of the order parameters for four different droplet configurations, the point $(x_0, y_0) = (L/2, 0)$ indicates the coordinates of the circle center with $L$ being size of square domain, $R$ is the droplet radius, and $\epsilon$ is the interface thickness. }	
	\begin{tabular}{cc}
		\hline
		Case & Initialization of order parameters                                                                                                                                                                                                                                                                                                                                                                                                                                                                                                                                                                                                                                                                                                     \\ \hline
		Janus          & {\color[HTML]{000000} $\begin{aligned} & \phi_1(x, y)=\left(\frac{1}{2}+\frac{1}{2} \tanh \frac{R-\sqrt{\left(x-x_0\right)^2+\left(y-y_0\right)^2}}{\epsilon / 2}\right)\left(\frac{1}{2}-\frac{1}{2} \tanh \frac{2\left(x-x_0\right)}{\epsilon}\right), \\ & \phi_2(x, y)=\left(\frac{1}{2}+\frac{1}{2} \tanh \frac{R-\sqrt{\left(x-x_0\right)^2+\left(y-y_0\right)^2}}{\epsilon / 2}\right)\left(\frac{1}{2}+\frac{1}{2} \tanh \frac{2\left(x-x_0\right)}{\epsilon}\right), \\ & \phi_3(x, y)=1-\phi_1(x, y)-\phi_2(x, y).\end{aligned}$}                                                                                                                                                                                    \\ \hline
		Collar         & $\begin{aligned}  \phi_2(x, y)=&\left(\frac{1}{2}+\frac{1}{2} \tanh \frac{R-\sqrt{\left(x-x_0\right)^2+\left(y-y_0\right)^2}}{\epsilon / 2}\right)\left(\frac{1}{2}-\frac{1}{2} \tanh \frac{x+\sqrt{2-\pi / 4} R-x_0}{\epsilon / 2}\right)+ \\ & \left(\frac{1}{2}+\frac{1}{2} \tanh \frac{R-\sqrt{\left(x-\sqrt{2-\pi / 4} R-x_0\right)^2+\left(y-y_0\right)^2}}{\epsilon / 2}\right)\left(\frac{1}{2}+\frac{1}{2} \tanh \frac{x-\sqrt{2-\pi / 4} R-x_0}{\epsilon / 2}\right), \\  \phi_1(x, y)=&\left[1-\phi_2(x, y)\right]\left(\frac{1}{2}+\frac{1}{2} \tanh \frac{\sqrt{2-\pi / 4} R-\sqrt{\left(x-x_0\right)^2+\left(y-y_0\right)^2}}{\epsilon / 2}\right), \\  \phi_3(x, y)=&1-\phi_1(x, y)-\phi_2(x, y).\end{aligned}$ \\ \hline
		Lens           & {\color[HTML]{000000} $\begin{aligned} & \phi_2(x, y)=\left(\frac{1}{2}+\frac{1}{2} \tanh \frac{\sqrt{2} R / 2-\sqrt{\left(x-x_0\right)^2+\left(y-y_0\right)^2}}{\epsilon / 2}\right), \\ & \phi_1(x, y)=\left(\frac{1}{2}+\frac{1}{2} \tanh \frac{R / 2-\sqrt{\left(x-x_0\right)^2+\left(y-y_0-\sqrt{2} R / 2-R / 2\right)^2}}{\epsilon / 2}\right), \\ & \phi_3(x, y)=1-\phi_1(x, y)-\phi_2(x, y).\end{aligned}$}                                                                                                                                                                                                                                                                                                            \\ \hline
		Encapsulated   & $\begin{aligned} & \phi_1(x, y)=\left(\frac{1}{2}+\frac{1}{2} \tanh \frac{\sqrt{2} R / 2-\sqrt{\left(x-x_0\right)^2+\left(y-y_0\right)^2}}{\epsilon / 2}\right), \\ & \phi_2(x, y)=\left[1-\phi_1(x, y)\right]\left(\frac{1}{2}+\frac{1}{2} \tanh \frac{R-\sqrt{\left(x-x_0\right)^2+\left(y-y_0\right)^2}}{\epsilon / 2}\right), \\ & \phi_3(x, y)=1-\phi_1(x, y)-\phi_2(x, y) .\end{aligned}$                                                                                                                                                                                                                                                                                                                                \\ \hline
	\end{tabular}
	\label{Tab33}
\end{table}

\begin{figure}[H]
	\centering
	\includegraphics[width=0.6\textwidth]{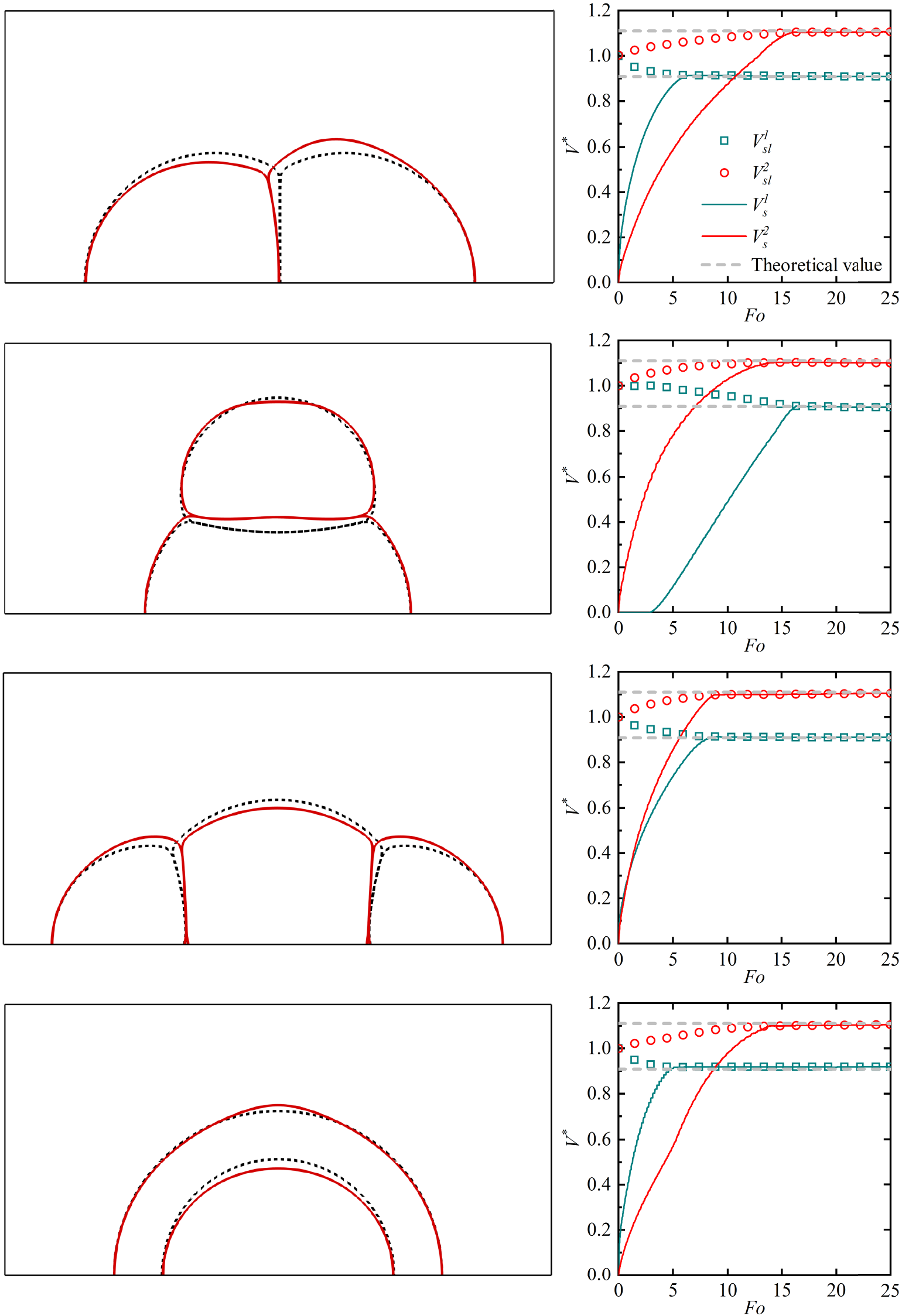}
	\put(-293, 400){(\textit{a})}
	\put(-293, 296){(\textit{b})}
	\put(-293, 193){(\textit{c})}
	\put(-293, 90){(\textit{d})}	
	\caption{ Phase interfaces of composite sessile droplets before and after solidification (left) and the corresponding evolutions of phase volumes (right). The configurations include: (a) Janus configuration, (b) collar configuration, (c) lens configuration and (d) encapsulated configuration. The black dashed line denotes the initial phase interface, while the red solid line represents the phase interface contour after complete freezing. }
	\label{fig11}
\end{figure}
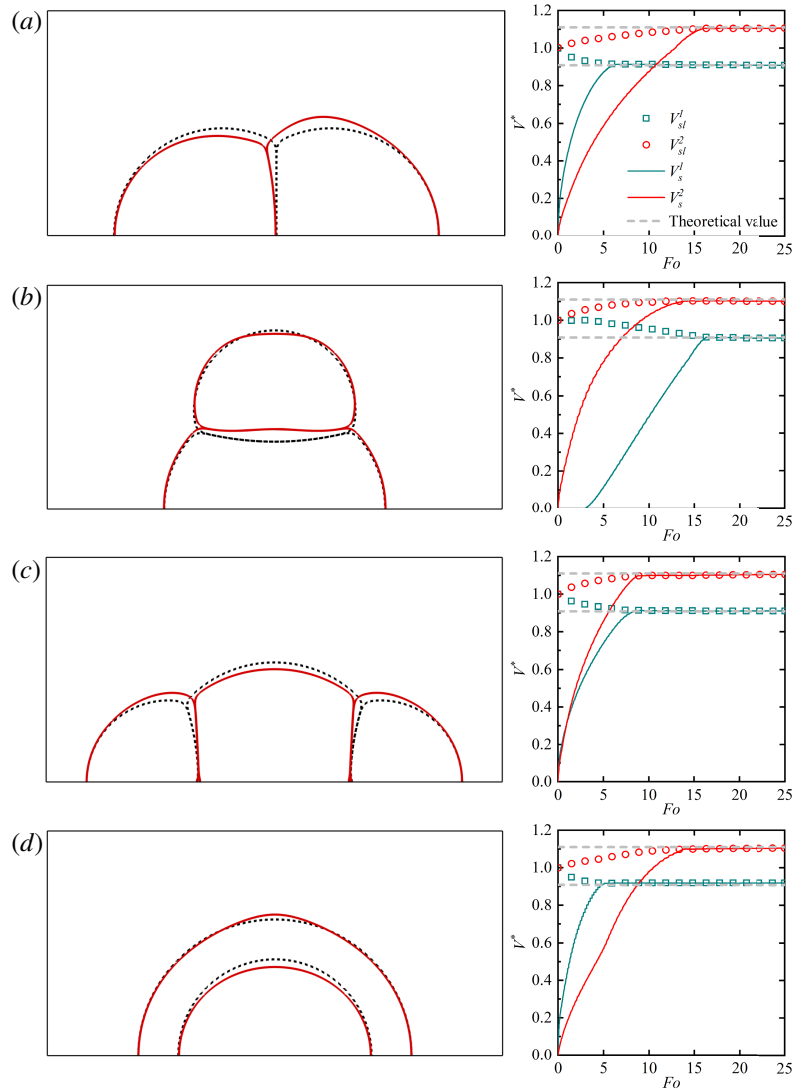

Fig. \ref{fig11} presents some comparisons of the phase interfaces before and after freezing (left) and the corresponding volume evolution of each phase during the freezing process (right) for four different configurations. During the freezing process, the volume of droplet 1 shrinks, while the volume of droplet 2 expands, resulting in deformation of the phase interface. Compared to other configurations, lens-shaped droplet freezes more quickly due to the larger solid-liquid contact area and shorter heat conduction path, which facilitate more efficient heat transfer. Such behavior aligns with the freezing characteristics of pure droplet. Furthermore, the total mass of the composite droplet is strictly conserved throughout the phase change process. These results demonstrate that the current diffuse-interface method is suitable for the \textit{N}-phase wetting fluids with phase change.

\subsection{Freezing with impurities}

The interaction between the freezing front and dispersed impurities (such as bubbles, droplets, and solid particles) is ubiquitous in both natural phenomena and engineering applications, such as sea ice formation and additive manufacturing \cite{HuerreARFM2024, DuNRP2024, ZhangNC2020}. When the freezing front approaches suspended bubbles or immiscible droplets suspended in the liquid phase, it can engulf them into the growing solid region or repel them \cite{ParkJFM2006, TyagiSR2021, BuurenPRL2024}. We now focus on the complex case of the freezing process in a liquid pool containing impurities and immiscible droplets.  

\begin{figure}[H]
	\centering
	\includegraphics[width=0.9\textwidth]{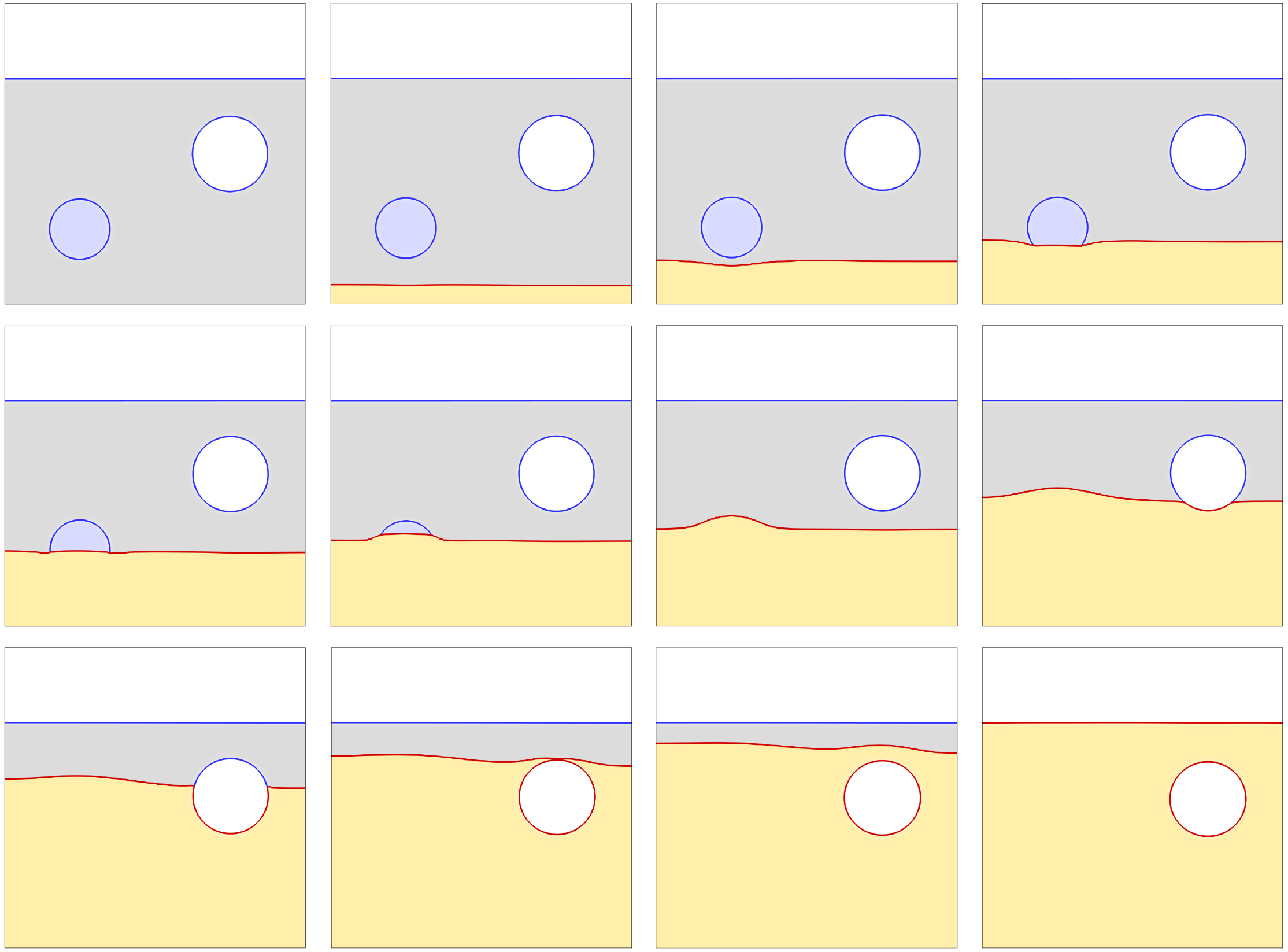}
	\put(-419, 303){$Fo=0.0$}
	\put(-312, 303){$Fo=0.03$}
	\put(-206, 303){$Fo=0.13$}
	\put(-99, 303){$Fo=0.25$}
	\put(-419, 197.4){$Fo=0.34$}
	\put(-312, 197.4){$Fo=0.42$}
	\put(-206, 197.4){$Fo=0.51$}
	\put(-99, 197.4){$Fo=0.72$}
	\put(-419, 92){$Fo=1.02$}
	\put(-312, 92){$Fo=1.23$}
	\put(-206, 92){$Fo=1.36$}
	\put(-99, 92){$Fo=1.71$}
	\caption{ Phase distribution and interface evolution during the solidification of a multiphase liquid pool containing impurities. The blue, gray, and white regions correspond to the droplets composed of fluid 1, fluid 2, and the gas phase, respectively. The yellow region represents the solid phase formed during the freezing process. Blue and solid lines denote the phase interface and advancing solidification front. }
	\label{fig12}
\end{figure}
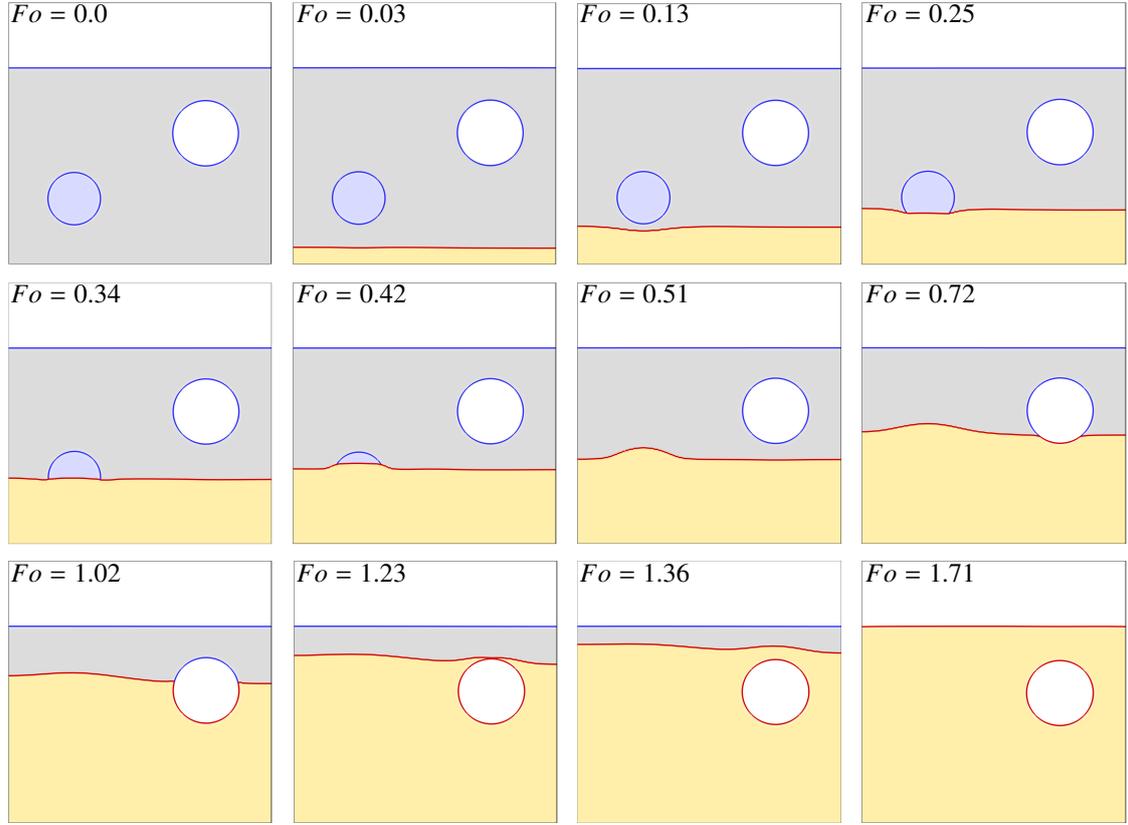

\begin{figure}[H]
	\centering
	\includegraphics[scale=0.4]{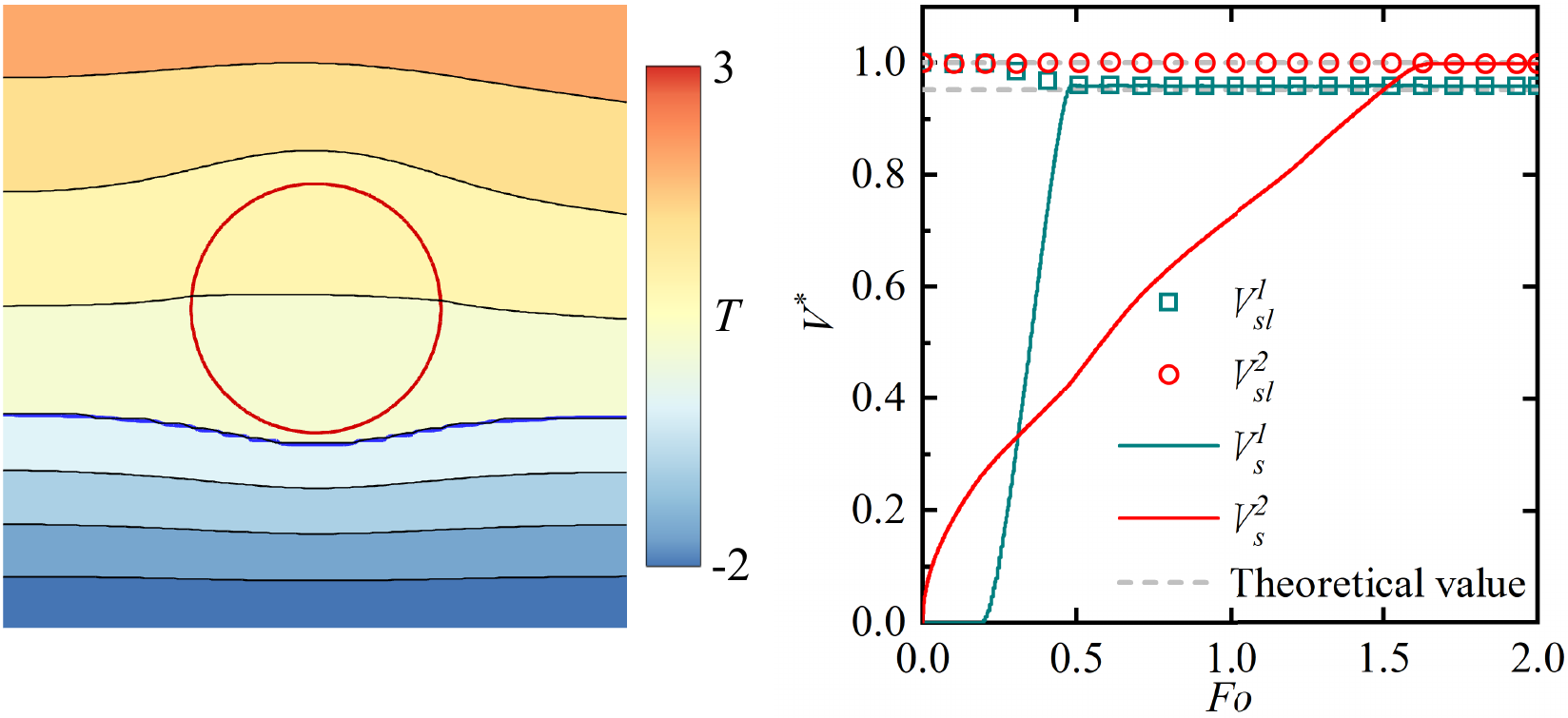}
	\put(-330 ,137){(\textit{a})}
	\put(-160 ,137){(\textit{b})}
	\caption{ (a) Isotherm distribution as the freezing front approaches droplet 1 at $Fo=0.17$. The red line indicates the interface of droplet 1, and the blue solid line represents the freezing front. (b) The dimensionless volume $V^*$ changes of the phases during the freezing process. }
	\label{fig13}
\end{figure}

In the computational domain $L \times L$, the lower region with $3L/4$ is filled with fluid 2, while the upper region with $L/4$ is occupied by fluid 3. The base fluid 2 contains a suspended droplet composed of fluid 1 and a bubble composed of fluid 3. For the flow field, the no-slip boundary condition is used for all boundaries, while for the temperature field, a low temperature is imposed on the bottom substrate to drive solidification, and the other boundaries are adiabatic. It should be noted that for this problem, only the droplet and fluid 2 undergo phase change. Fig. \ref{fig12} shows the dynamic evolution of the phase interface during the freezing process. It is found that in the regions far from the impurities, the freezing front initially remains nearly parallel to the bottom wall. However, as the freezing front approaches the droplet, the morphology is changed significantly, curving downward and deviating from the droplet. This morphological change is consistent with previous experimental observation of freezing front-droplet interaction. Actually, the freezing front morphology is determined by the freezing isotherm ($T = T_m$), which is strongly dependent on the distribution of thermal diffusivity. When the thermal diffusivity of the impurity is higher than that of the surrounding base fluid 2, the heat tends to preferentially conduct through the impurity. As a result, the isotherms beneath the impurity shift [see Fig. \ref{fig13}(a)], which bend the melting isotherm downward and creates a depressed freezing front. This phenomenon is consistent with the experimental results \cite{TyagiSR2021, BuurenPRL2024}. As the freezing progresses, the droplet with higher thermal diffusivity is gradually engulfed by the freezing front, and the morphology of the freezing front gradually changes from concave to convex pattern due to the high thermal diffusivity of the droplet, which accelerates the heat transfer in and around the droplet. Eventually, the bubbles composed of fluid 3 are captured by the freezing front, creating pore defect in the solid phase.

Furthermore, Fig. \ref{fig13}(b) presents the temporal evolutions of the solid-liquid mixture volumes of fluids 1 and 2 during the solidification process, along with the solid-phase volumes. Initially, the droplet composed of fluid 1 remains in the liquid state, and its solid-phase volume ($V_1^s$) is zero. With the increase of time, the droplet undergoes volume contraction due to $\rho_1^s /\rho_1^l =1.05>1$, while the base fluid 2 solidifies with the volume expansion owing to $\rho_2^s /\rho_2^l =0.95<1$. The results also indicates that the developed diffuse-interface method can effectively capture the contrasting volume responses during phase change: shrinkage of the droplet and expansion of fluid 2.

\section{Conclusions}
\label{sec5}
We propose a diffuse-interface model to study the \textit{N}-phase flows with solid–liquid phase change. The phase change is modeled by an enthalpy-based approach, enabling the accurate modeling of quasi-equilibrium isotropic solidification without protrusions or dendritic structures, and simultaneously, the dynamics of immiscible \textit{N}-phase flows are governed by a conservative AC equation. Through coupling the phase-field order parameters with the solid fraction, the model can effectively capture the evolutions of phase interfaces between different phases during the solidification process. The diffuse-interface model can account for volume change caused by density differences between the solid and liquid phases, while ensuring mass conservation of the phase-change materials. It also satisfies consistent condition, allowing it to rigorously degenerate to both the incompressible \textit{N}-phase conservative phase-field model and the classical enthalpy method for solid–liquid phase change. To solve the present diffuse-interface model, a diffuse-interface LB method is further developed, and three benchmark tests are also conducted to validate the method, including the film freezing, single droplet freezing, and compound droplet freezing. The results show that the diffuse-interface LB method not only is suitable for the \textit{N}-phase flows with phase change, but also has a good numerical performance in preserving mass conservation of phase-change materials. In addition, the method can also capture the distinct features of liquid phases during the freezing process, such as the formation of tip and plateau structures in the frozen droplets. Furthermore, the freezing dynamics of \textit{N}-phase system containing impurities is investigated, and the numerical results demonstrate that the difference in thermal diffusivity has an important influence on the morphology of the freezing front, while the model consistently enforces mass conservation even for complex phase-change scenarios. While the proposed model effectively captures phase change in N-phase systems, it relies on the assumption of a uniform melting temperature for all species, which remains to be addressed in future work.

\section*{Acknowledgments}
The computation was completed in the HPC Platform of Huazhong University of Science and Technology. This work was financially supported by the National Natural Science Foundation of China (Grant Nos. 12501599 and 123B2018), the Postdoctoral Fellowship Program of CPSF (Grant No. GZB20250714), China Postdoctoral Science Foundation (Grant No. 2025M773077), the Open Research Fund of Stake Key Laboratory of Mesoscience and Engineering (Grant No. MESO-25-D04) and the Interdisciplinary Research Program of Hust (Grant No. 2024JCYJ001).


\end{document}